\documentclass[prb,twocolumn,amsmath,amssymb,floats,showpacs,superscriptaddress]{revtex4-1}
\usepackage{graphicx}
\usepackage{bm}
\usepackage{amsmath}
\usepackage{amsfonts}
\usepackage{amsbsy}
\usepackage{verbatim}
\usepackage{longtable}
\usepackage{color}
\usepackage{subfigure}

\newcommand\abs[1]{\lvert#1\rvert}

\begin{document}

\title{Anomalous Josephson current through a driven double quantum dot}

\author{Carlos Ortega-Taberner}
\affiliation{Department of Physics, Stockholm University, AlbaNova University Center, SE-106 91 Stockholm, Sweden}
\affiliation{Nordita, KTH Royal Institute of Technology and Stockholm University, SE-106 91 Stockholm, Sweden}

\author{Antti-Pekka Jauho}
\affiliation{Center for Nanostructured Graphene, Department of Physics,
Technical University of Denmark, DK-2800 Kongens Lyngby, Denmark}

\author{Jens Paaske}
\affiliation{Center for Quantum Devices, Niels Bohr Institute, University of Copenhagen, DK-2100 Copenhagen \O, Denmark}

\date{\today}
\begin{abstract}
Josephson junctions based on quantum dots offer a convenient tunability by means of local gates. Here we analyze a Josephson junction based on a serial double quantum dot in which the two dots are individually gated by phase-shifted microwave tones of equal frequency. We calculate the time-averaged current across the junction and determine how the phase shift between the drives modifies the current-phase relation of the junction. Breaking particle-hole symmetry on the dots is found to give rise to a finite average anomalous Josephson current with phase bias between the superconductors fixed to zero. This microwave gated weak link thus realizes a tunable "Floquet $\varphi_{0}$-junction" with maximum critical current achieved for driving frequencies slightly off-resonance with the energy cost of exciting a sub-gap state on each dot. We provide numerical results supported by an analytical analysis for infinite superconducting gap and weak inter-dot coupling. We identify an interaction driven $0-\pi$ transition of anomalous Josephson current as a function of driving phase difference. Finally, we show that this junction can be tuned so as to provide for complete rectification of the time-averaged Josephson current phase relation.

\end{abstract}

\pacs{72.10.Fk, 74.45.+c, 73.63.Kv, 74.50.+r}
\maketitle

\section{Introduction}

The Josephson junction (JJ) has become a ubiquitous device serving in a wide range of applications, including the superconducting qubits which have lead to impressive advances in quantum computing during the past two decades~\cite{Makhlin2001May, Blais2004Jun, Clarke2008Jun, Girvin2011, Kjaergaard2020Mar}. The weak link coupling the two superconductors can either be a plain insulating tunnel barrier, or it may exhibit internal structure like a normal region, a quantum point contact, a magnetic tunnel barrier, or a quantum dot (QD), which all host sub-gap states which may influence strongly the current phase relation (CPR) of the junction~\cite{Kulik1966Sep, Shiba1969Jan, Beenakker1991Dec, Glazman1989, Rozhkov1999Mar, Martin-Rodero2011Dec, Meden2019Feb}. In this way, electrically gateable links like quantum dots or semiconductors offer a certain tunability of the JJ characteristics~\cite{vanDam2006Aug, Delagrange2016May, vanWoerkom2017, Bouman2020Dec}, a feature which has been employed in the design of a hybrid gatemon~\cite{Larsen2015Sep, Casparis2016Apr, Casparis2018Oct}, adding gate control to the superconducting transmon qubit~\cite{Koch2007Oct, DiCarlo2009Jul}, which has already demonstrated its efficiency in solid state quantum computing~\cite{Kelly2015Mar, Kandala2017Sep, Neill2018Apr}.

Whereas normal Josephson junctions carry no current at zero phase bias, $\varphi_{sc}=\varphi_{L}-\varphi_{R}$, a weak link which breaks both time-reversal and chiral symmetry may carry an {\it anomalous Josephson current} between two superconductors maintained at zero phase bias~\cite{Zazunov2009}. A number of proposals~\cite{Geshkenbein1986, Buzdin2003, Reynoso2008, Zazunov2009, Tanaka2009, Liu2010, Goldobin2011, Brunetti2013, Alidoust2013, Yokoyama2014, Campagnano2015, Bergeret2015, Dolcini2015, Alidoust2020Apr} have been made for such $\varphi_{0}$ junctions with an anomalous Josephson current, $I(\varphi_{sc})=I_{C}\sin(\varphi_{sc}+\varphi_0)$, at least two of which have already been realized experimentally~\cite{Sickinger2012, Szombati2016}. Of particular relevance to the present work is the proposal by Zazunov et al.~\cite{Zazunov2009} to use a multi-orbital QD with inter-orbital (spin-orbit) tunnelling and an external field. With such a link in the JJ, traversing electrons pick up different phases, depending on the tunnelling direction, giving rise to an anomalous Josephson current. This proposal has since been realized in an experiment by Szombati et al.~\cite{Szombati2016}, using an InSb-wire QD contacted by superconducting NbTiN leads.

\begin{figure}[t]
\includegraphics[width = .9\columnwidth]{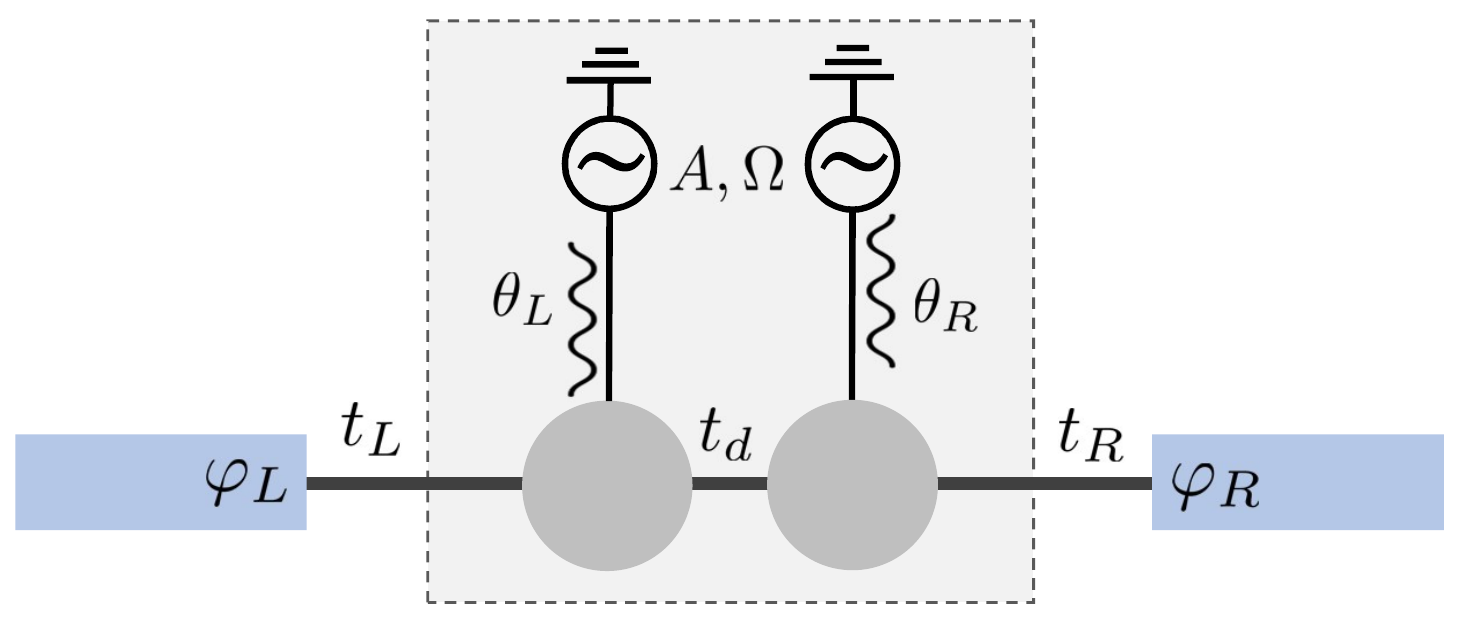}
\caption{Sketch of a Josephson junction with a structured weak link (gray region) based on a driven double quantum dot. The superconductors (blue) are maintained at a fixed phase-bias $\varphi_{sc}=\varphi_{L}-\varphi_{R}$, and the weak link is driven by two microwave gates with same amplitude and frequency, $A, \Omega$, shifted in phase by $\theta_{d}=\theta_{L}-\theta_{R}$. The internal, and the two external tunnelling amplitudes are denoted by $t_{d}$, $t_{L}$ and $t_{R}$, respectively.}
\label{fig:setup_2qd}
\end{figure}
Here, we propose a nonequilibrium version of the multi-orbital QD considered in Ref.~\onlinecite{Zazunov2009}, based on the device illustrated in Fig.~\ref{fig:setup_2qd}. In this Josephson junction, the two superconductors are coupled by a serial double quantum dot (DQD) where the two dots are driven by individual AC gate voltages with a common amplitude, $A$, and microwave frequency, $\Omega$. This endows each of the QDs with Floquet sidebands, which play the roles of the additional spin-orbit coupled orbitals in Ref.~\onlinecite{Zazunov2009}. As we demonstrate below, the phase difference between the two drive voltages, $\theta_d=\theta_{L}-\theta_{R}$,  can have a strong influence on the JJ CPR, and with QD levels tuned away from particle-hole symmetry it gives rise to anomalous current, which in the limit of weak tunnel couplings reduces to a simple $\varphi_{0}$ junction, with $\varphi_{0}=\theta_d$. Since the time-averaged critical current is maximized when the microwave frequency is close to the energy for exciting both of the sub-gap states induced in the two proximitized quantum dots, this device comprises a non-adiabatic Cooper pair pump, or more aptly a "Floquet $\varphi_{0}$ junction".

The undriven DQD Josephson junction with individual gating of the two dots has already been realized experimentally~\cite{Su2017Sep, EstradaSaldana2018Dec, Saldana2020Nov, Bouman2020Dec}, and understood to constitute a strongly correlated transport problem involving the formation of sub-gap states, which depend strongly on the charge configuration of the (Coulomb blockaded) dots~\cite{Bauer2007, Meng2009, Droste2012, Brunetti2013Oct, Kirsanskas2015Dec, Meden2019Feb, Kadlecova2019Apr, Pokorny2020Jan, Bouman2020Dec}. Here we circumvent a number of these complications by replacing each of the dots with a non-interacting resonant level. Whereas this would clearly be a poor description of Coulomb blockaded QDs in many other respects, the two models do share the crucial features of the mechanism we wish to illustrate, namely the presence of sub-gap states with a strong gate dependence.
As a weak link for a JJ, the resonant level model behaves much like a quantum point contact (QPC) with a CPR which reflects the phase dispersion of the sub-gap Andreev bound states (ABS)~\cite{Beenakker1991Dec, Bauer2007, Haller2014}. A JJ based on a Coulomb blockaded QD, however, is known to exhibit a transition from a $\varphi_{0}=\pi$ to $\varphi_{0}=0$ phase~\cite{Glazman1989, Rozhkov1999Mar, Delagrange2016May, Kirsanskas2015Dec, Meden2019Feb, Kadlecova2019Apr}, and the results presented below are therefore of greater relevance for a realistic Coulomb blockaded QD in its $\varphi_{0}=0$ phase stabilized for strong tunnel couplings~\cite{Hermansen2022Feb}, or for a long normal junction with a finite dwell time~\cite{Kurilovich2021Nov}. 

Dating back to the seminal work on photon assisted tunnelling by Tien and Gordon~\cite{Tien1963}, the problem of non-adiabatically (microwave) driven Josephson junctions has been expanded to include also junctions with QPC, QD, DQD or magnetic adatom weak links~\cite{Grifoni1998Oct, Platero2004, Bergeret2010, Bergeret2011, Kos2013May, Bretheau2014Oct, Olivares2014Mar, Venitucci2018, Gonzalez2020Jul, Baran2021May}. Experimentally, the ABS in such junctions have been measured and manipulated using microwave spectroscopy~\cite{Bretheau2013Jul, Janvier2015Sep, Blais2020May, Peters2020Dec, Canadas2021Dec, Fatemi2021Dec}, and these techniques are by now becoming widely available. Recently, Venitucci et al.~\cite{Venitucci2018} demonstrated that phase shifted microwave voltages applied to each of the superconductors in a JJ with a single resonant level as the weak link can give rise to photon assisted Cooper pair transfer and a tunable $\varphi_0$-junction. Similarly, Soori et al.~\cite{Soori2020Jun} have studied a finite-size tight-binding model of an SNS junction and found that a phase shifted drive on the two sites comprising their normal region leads to anomalous Josephson current. The model studied here is similar in spirit but not equivalent to these two studies, and our main focus is the anomalous Josephson current and the modified CPR for the JJ driven at resonance with the sub-gap states.

The paper is organized as follows. In Sec.~\ref{sec:model} we present the model. In Sec.~\ref{sec:Greens} we define the relevant Nambu-Floquet-Keldysh Green functions and provide an expression for the time-averaged current to be calculated. In Sec.~\ref{infgap} we study the limit of infinite gap, in which the main effect of the $\varphi_{0}$-junction can be established analytically in the limit of weak inter-dot tunnel coupling. Sec.~\ref{sec:numerics} contains the numerical results for the current and the CPR for the driven junction. Finally, the results are discussed in Sec.~\ref{sec:discuss}. Appendix~\ref{app} provides a supplementary analysis for the infinite-gap limit using Floquet theory, which allows us to also investigate the effects of local Coulomb interactions, and to confirm the rectification of the time-averaged supercurrent.

\section{The model}\label{sec:model}

We consider a non-interacting serial double quantum dot with on-site energies modulated by individual AC gate voltages and coupled to two (Left/Right) superconducting leads (cf. Fig.~\ref{fig:setup_2qd}). The Hamiltonian reads
\begin{equation}
H(t)=\sum_{\alpha=L,R}H_{\rm sc,\alpha}+H_{\rm d}(t)+H_{\rm t},
\end{equation}
with superconducting leads described by BCS Hamiltonians
\begin{align}
H_{\rm sc,\alpha}\!=\!\sum_{{\bf k},\sigma}\left[
\xi_{\alpha {\bf k}}c^{\dagger}_{\alpha {\bf k} \sigma}c_{\alpha {\bf k} \sigma}
+\left(\Delta e^{i\varphi_{\alpha}} c_{\alpha {\bf k} \uparrow} c_{\alpha-{\bf k}\downarrow}+\text{h.c.}\right)\right],
\end{align}
for $\alpha = L,R$. The two leads are kept at the same chemical potential and are assumed to have the same gap magnitude, $\Delta>0$, with different phases, $\varphi_{L,R}=\pm\varphi_{sc}/2$. Both leads are represented by a featureless band-structure near a common chemical potential, i.e. $\xi_{\alpha {\bf k}}=\varepsilon_{\alpha {\bf k}}-\mu$, corresponding to a common density of states, $\nu_{F}$, near the Fermi level. The time-dependent Hamiltonian of the double quantum dot system reads
\begin{align}
H_{\rm d}(t)=&\!\!\!\!\sum_{\sigma;\alpha,\alpha'\in L,R}\!\!\!d^{\dagger}_{\alpha \sigma}\left[\varepsilon_{d\alpha}(t)\tau^{0}_{\alpha \alpha'} + t_{d}\tau^{x}_{\alpha\alpha'}\right] d_{\alpha' \sigma},
\end{align}
with individual AC gate voltages given as $\varepsilon_{d\alpha}(t)=\varepsilon_{d}+A\cos(\Omega t+\theta_\alpha)$, in terms of common (time) average energies, $\varepsilon_d$, driving amplitudes, $A$, frequencies, $\Omega$, and two independent phase constants, $\theta_\alpha$. Here, $\tau^{i}$ denotes the $i$'th Pauli matrix, $\tau^{0}$ the Kronecker delta and $t_d$ is the inter-dot tunneling amplitude. The tunneling Hamiltonian reads
\begin{align}
H_{\rm t}=\sum_{{\bf k},\sigma,\alpha=L,R}t_{\alpha } c_{\alpha {\bf k} \sigma}^{\dagger}d^{}_{\alpha \sigma}+\text{h.c.}.
\end{align}
Written in terms of Nambu spinors, $\psi^{\dagger}_{\alpha {\bf k}} = (c_{\alpha {\bf k} \uparrow}^{\dagger}  \,, c_{\alpha -{\bf k} \downarrow})$ and $\phi^{\dagger}_{\alpha} = (d_{\alpha \uparrow}^{\dagger} \,, d_{\alpha \downarrow})$, the full Hamiltonian reads
\begin{align}\label{eq:Hamt}
H(t)=&\sum_{\alpha {\bf k}}\psi^{\dagger}_{\alpha {\bf k}}
(\xi_{\alpha {\bf k}}\sigma_{z}-\Delta \sigma_{x}) \psi_{\alpha {\bf k} }\nonumber\\
&+ \sum_{\alpha \alpha'}\phi_{\alpha}^{\dagger}\left(\varepsilon_{d\alpha}(t)\tau^0_{\alpha \alpha'}+t_d\tau^x_{\alpha \alpha'}\right)\sigma_z\phi_{\alpha'} \nonumber\\
&+\sum_{\alpha {\bf k}} \left( \psi^{\dagger}_{\alpha {\bf k} }\mathcal{T}_{\alpha}\phi_{\alpha} +\phi^{\dagger}_{\alpha}\mathcal{T}_{\alpha}^{*}\psi_{\alpha {\bf k} }\right),
\end{align}
where the phase of the superconducting leads has been gauged into the tunneling matrix, $\mathcal{T}_{\alpha} = t_{\alpha}\sigma_z e^{i\sigma_z \varphi_{\alpha}/2}$. For simplicity, we assume below that tunneling amplitudes to the leads are real and equal, i.e. $t_L=t_R\equiv t$.

As discussed in the introduction, we neglect the charging energies of both quantum dots altogether and consider this resonant level model as an effective model for a proximitized QD with a doublet sub-gap state.

\section{Keldysh Floquet Green functions}~\label{sec:Greens}

To calculate the current through the ac-driven device, we employ the non-equilibrium Green function technique~\cite{Keldysh1964, Rammer1986Apr, Haug2008}. Dealing with a harmonic drive, it is convenient to use Floquet Keldysh Green functions~\cite{Faisal1989Feb, Tsuji2008}, which offers a representation of the two-time Green functions, which, besides being convenient for numerical calculations, allows for some degree of physical interpretation of the elementary transport process in terms of Floquet side bands. The time-dependent current out of lead $\alpha$ for this driven junction is found as~\cite{Haug2008}
\begin{align}\label{eq:current_keldysh}
J_{\alpha}(t) =& 2 \text{Tr}\left\{\sigma_{z}\text{Re}  \int dt_1 \, \left[ G^{R}_{d,\alpha\alpha}(t,t_1)\Sigma^< _{\alpha}(t_1,t) \right.\right.\\
&\hspace*{3cm}+\left.\left.G^<_{d,\alpha\alpha}(t,t_1)\Sigma^{A}_{\alpha}(t_1,t)\right] \vphantom{\int} \right\},\nonumber
\end{align}
where the trace is taken in Nambu space, and with Nambu/lead-matrix Green functions for the quantum dots defined as
\begin{align}
G^{R,A}_{\alpha\eta,\alpha'\eta'}(t,t')=&\mp i\theta(\pm t \mp t') \langle \{\phi_{\alpha\eta}(t),\phi^{\dagger}_{\alpha'\eta'}(t')\}\rangle\\
G^{<}_{\alpha\eta,\alpha'\eta'}(t,t')=&\,i\langle\phi^{\dagger}_{\alpha'\eta'}(t')\phi^{}_{\alpha\eta}(t)\rangle,\\
G^{>}_{\alpha\eta,\alpha\eta'}(t,t')=&\,-i\langle\phi^{}_{\alpha\eta}(t)\phi^{\dagger}_{\alpha'\eta'}(t')\rangle,\nonumber
\end{align}
with self-energies, which are exact to second order in dot-lead tunnelling,
\begin{align}\label{eq:lead_gf}
\Sigma^{R,A,<}_{\alpha}(t)=\sum_{k}
\mathcal{T}^{\ast}_{\alpha}g^{R,A,<}_{\alpha k}(t)\mathcal{T}_{\alpha} ,
\end{align}
where $g_{\alpha k}$ denotes the Nambu Green function in lead $\alpha$. From this self-energy, the dot Green functions can be found by solving the steady-state Dyson equations,
\begin{align}
\!G^{R/A}(t,t')\!=&\,G^{R/A(0)}(t,t_1)+\int\!\!dt_{1}dt_{2} G^{R/A(0)}(t,t_1)\nonumber\\
&\times\Sigma^{R/A}(t_1-t_2)G^{R/A}(t_2,t'),\label{eq:GRDyson}\\
G^<(t,t')\!=& \int\!\!dt_{1}dt_{2}G^R(t,t_1)\Sigma^<(t_1-t_2)\nonumber\\
&\times G^A(t_2,t'),\label{eq:GLDyson}
\end{align}
with matrix products between Green functions implied.

With a periodic drive, it is convenient to transform these two-time Green functions into Floquet matrices ~\cite{Tsuji2008}
\begin{align}
\!\!\!O_{nm}(\omega)=\!\int_{-\infty}^{\infty}\!\!\!\!dt'\frac{1}{T}\!\int_{0}^{T}\!\!dt\, e^{i(\omega+n\Omega)t-i(\omega+m\Omega)t'}O(t,t'),\label{eq:FLtransf}
\end{align}
defined with $\omega\in]-\Omega/2, \Omega/2]$. This transformation presumes that the Green functions are periodic in both time arguments, with the driving period $T=2\pi/\Omega$, and thereby rests on the assumption that the system has reached a nonequilibrium steady state (NESS). In this way, the time-averaged current may be found as 
\begin{align}
J^0_{\alpha} =& \frac{1}{T}\int_0^T dt \, J_\alpha (t)\\ =&2\text{Tr}\left\{\sigma_{z}\text{Re} \int _{-\Omega/2}^{\Omega/2}d\omega \, \left[ G^R_{d,\alpha \alpha}(\omega) \Sigma^< _{\alpha}(\omega)\right.\right.\\
&\left.\left.\hspace*{42mm}+G^<_{d,\alpha \alpha}(\omega) \Sigma^{A}_{\alpha}(\omega)  \right] \right\}.\nonumber
\end{align}
This is the zeroth Floquet component of the current. Here, the Green functions and self-energies are matrices in Nambu, dot, and Floquet space and the trace is performed over all these spaces. The components of the self-energy in dot, and Floquet space are given by
\begin{align}\label{eq:selfen}
\Sigma^{R,A,<}_{\alpha,nm}(\omega)= &
\mathcal{T}^{\ast}_{\alpha}g^{R,A,<}_{\alpha}(\omega+n\Omega)\mathcal{T}_{\alpha} \delta_{nm},
\end{align}
where the momentum-summed lead Nambu Green functions are given explicitly as
\begin{align}
g_{\alpha}^{R,A}(\omega)
=&\pi\nu_{F}\frac{-(\omega\pm i 0_+)\sigma^{0}+\Delta \sigma^{x}}{\sqrt{\Delta^2-(\omega\pm i 0_+)^2}},\label{eq:gR}\\
g_{\alpha}^{<}(\omega)=&n_{F}(\omega)\left(g_{\alpha}^{A}(\omega)-g_{\alpha}^{R}(\omega)\right),
\end{align}
where $n_F$ denotes the Fermi function. Henceforth, temperature is assumed to be zero.

Finally, using the Dyson equation~\eqref{eq:GRDyson}, the retarded double-dot Green function is found by inverting the following infinite dimensional Floquet matrix of $4\times 4$ matrices in Nambu-dot space:
\begin{align}\label{eq:GRinv}
\left(G^{R,A}_{d}(\omega)\right)^{-1}_{\alpha \alpha',nm}=&
\left\{-t_{d}\sigma^{z}\tau^{x}_{\alpha\alpha'}\right.\\
&\hspace*{-2.6cm}\left.+\left[(\omega+n\Omega)\sigma^{0}-\mathcal{T}_{\alpha }^* g^{R,A}_{\alpha}(\omega+n\Omega)\mathcal{T}_{\alpha}- \varepsilon_d\sigma^{z}\right]\tau^{0}_{\alpha\alpha'}\right\}\delta_{nm}\nonumber \\
&\hspace*{-2.6cm}-\frac{A}{2}\left(e^{-i \theta_{\alpha}/2}\delta_{n-m,1}+e^{i \theta_{\alpha}/2}\delta_{m-n,1}\right)\sigma^{z}\tau^{0}_{\alpha \alpha'}.\nonumber
\end{align}
From the resulting retarded and advanced Green functions, the lesser function is found from Eq.~\eqref{eq:GLDyson} by simple matrix multiplication.

\section{Infinite-gap limit with weak inter-dot tunnel coupling}~\label{infgap}

It is instructive to first consider the limit of an infinite superconducting gap. This prohibits quasiparticle tunnelling altogether and transport takes place only via Cooper pairs. In the infinite-gap limit, the retarded QD self-energy becomes
\begin{align}
\mathcal{T}_{\alpha }^* g^{R,A}_{\alpha}(\omega+n\Omega)\mathcal{T}_{\alpha}\approx
-\Gamma e^{-i\sigma_z \varphi_{\alpha}}\sigma_{x},
\end{align}
with $\Gamma=\pi\nu_F\abs{t}^2$, corresponding to an effective Hamiltonian describing a proximitized quantum dot with an induced superconducting gap of $\Gamma$:
\begin{align}\label{eq:infgh}
H_{\infty}(t)=&\sum_{\alpha=L,R}\phi^{\dagger}_{\alpha}
\left[\varepsilon_{d\alpha}(t)\sigma^{z}-\Gamma\sigma^{x}\right]\phi_{\alpha}
\\&\hspace*{1cm}+\phi^{\dagger}_{L}\mathcal{T}_{d}\phi^{}_{R}+\phi^{\dagger}_{R}\mathcal{T}_{d}^{\ast}\phi^{}_{L},\nonumber
\end{align}
with a matrix of tunnelling amplitudes given by $\mathcal{T}_{d,\eta\eta'}=|t_{d}|\sigma^{z}_{\eta\eta'}\exp(i\sigma^{z}_{\eta\eta'}\varphi_{sc}/2)$, with $\varphi_{sc}=\varphi_{L}-\varphi_{R}$.

In order to illustrate the basic microwave assisted Cooper pair transport mechanism within this infinite-gap model, we calculate here the weak-coupling tunnelling current to second order in the interdot coupling $t_{d}$, given by the perturbative expression~\cite{Ambegaokar1963Jun}:
\begin{align}\label{eq:td2curr}
I(t)&=2e|t_{d}|^{2}{\rm Re}\left[\int_{0}^{t}\!\!\!dt'
\sigma^{z}_{\eta'\eta'}e^{i(\sigma^{z}_{\eta'\eta'}-\sigma^{z}_{\eta\eta})\varphi_{sc}/2}\right.\\
&\hspace*{-6mm}\left(G^{>}_{L\eta,L\eta'}(t,t')G^{<}_{R\eta',R\eta}(t',t)
\!-\!G^{<}_{L\eta,L\eta'}(t,t')G^{>}_{R\eta',R\eta}(t',t)\right)\large]\nonumber
\end{align}
The driving enters this expression through the time-dependent correlation functions, $G^{<,>}_{\alpha\eta,\alpha\eta'}(t,t')$, describing the dynamics of the QD proximitized by lead $\alpha=L,R$.

The perturbative expression for the current requires the Green's functions for $t_{d}=0$, and in this case the Hamiltonian~\eqref{eq:infgh} describes two independent quantum dots.  It is readily diagonalized by the time-dependent Bogoliubov transformation (suppressing the QD index, $\alpha=L,R$):
\begin{align}
\chi_{\nu}=U_{\nu\eta}\phi_{\eta},
\hspace*{5mm}
\chi^{\dagger}_{\nu}=\phi^{\dagger}_{\eta}U^{-1}_{\eta\nu},
\end{align}
with Nambu spinors
\begin{align}
\chi=\left(
             \begin{array}{c}
               \gamma^{}_{\uparrow} \\
               \gamma^{\dagger}_{\downarrow} \\
             \end{array}
           \right),
\hspace*{5mm}
\phi=\left(     \begin{array}{c}
               d^{}_{\uparrow} \\
               d^{\dagger}_{\downarrow} \\
             \end{array}
           \right),
\end{align}
and time-dependent unitary transformation matrix
\begin{align}
U(t)=\left(
             \begin{array}{cc}
               u(t) & -v(t) \\
               v(t) & u(t) \\
             \end{array}
           \right),
\hspace*{5mm}
U^{-1}(t)=U^{T}(t),
\end{align}
with $E_{d}(t)=\sqrt{\varepsilon^{2}_{d}(t)+\Gamma^{2}}$ and real coherence factors given by
\begin{align}
u(t)=&\,\sqrt{(1+\varepsilon_{d}(t)/E_{d}(t))/2},\\
v(t)=&\,\sqrt{(1-\varepsilon_{d}(t)/E_{d}(t))/2}.\nonumber
\end{align}
Notice that we omit the $\alpha=L,R$ subscript for clarity since it only enters in the two different phase shifts, $\theta_{\alpha}$, and can readily be reinstalled. This transformation diagonalizes the Hamiltonian for each of the two different proximitized levels,
\begin{align}
H_{\infty}^{0}(t)=&\,\phi^{\dagger}\left[\varepsilon_{d}(t)\sigma^{z}-\Gamma\sigma^{x}\right]\phi=\,\chi^{\dagger}E_{d}(t)\sigma^{z}\chi^{},
\end{align}
and endows the quasiparticles with dynamics governed by the equation of motion,
\begin{align}
i\frac{d}{dt}\chi_{\nu}(t)=\left[
E_{d}(t)\sigma^{z}_{\nu\nu'}
+\frac{A\Omega\Gamma\sin(\Omega t+\theta)}{2E^{2}_{d}(t)}\sigma^{y}_{\nu\nu'}
\right]\chi_{\nu'}(t),
\end{align}
where the last term has been obtained as
\begin{align}\label{eq:mix}
-iU_{\nu\eta}(t)\left(\frac{d}{dt}U^{-1}_{\eta\nu'}(t)\right)
=\frac{A\Omega\Gamma\sin(\Omega t+\theta)}{2E^{2}_{d}(t)}\sigma^{y}_{\nu\nu'}.
\end{align}
The corresponding transformation of the correlation functions reads
\begin{align}\label{eq:Glg}
G^{<}_{\eta\eta'}(t,t')=&\,iU^{-1}_{\eta\nu}(t)U_{\nu'\eta'}(t')\langle\chi^{\dagger}_{\nu'}(t')\chi^{}_{\nu}(t)\rangle,\\
G^{>}_{\eta\eta'}(t,t')=&\,-iU^{-1}_{\eta\nu}(t)U_{\nu'\eta'}(t')\langle\chi^{}_{\nu}(t)\chi^{\dagger}_{\nu'}(t')\rangle.\nonumber
\end{align}

\begin{figure}[t]
\centering
\includegraphics[width = \columnwidth]{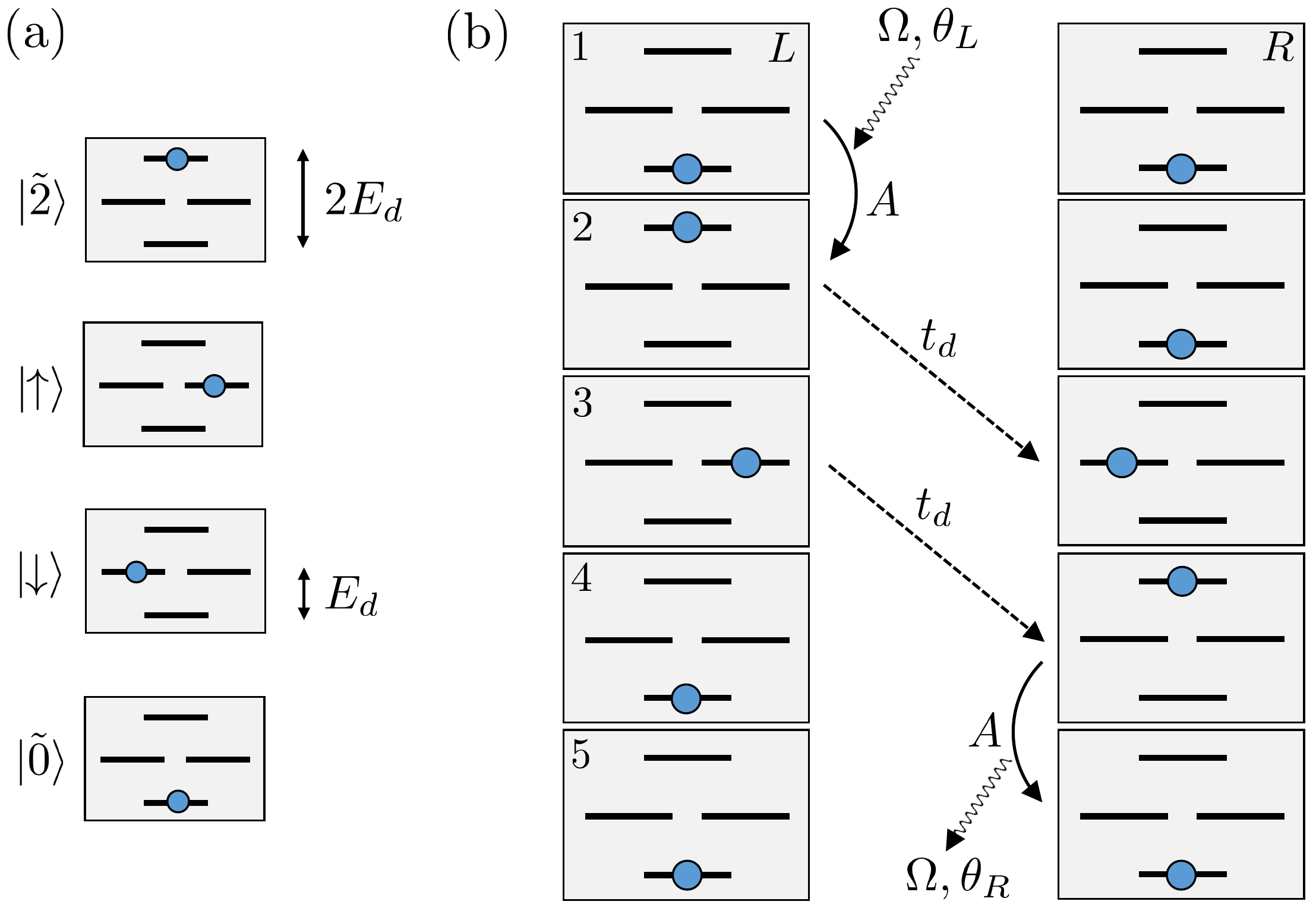}
\caption{
(a) A schematic of the four many-body eigenstates of a single proximitized undriven QD. The ground state has zero energy, the excited doublet has energy $E_{d}$, and the two-quasiparticle state has energy $2E_{d}$. (b)
Diagram illustrating the path of a Cooper pair through the driven DQD junction in progression from panel 1 through 5. Driving the microwave gates with $\Omega\sim 2E_{d}$ induces a near resonant transition from $|\tilde{0}\rangle$ to $|\tilde{2}\rangle$ in the left QD (1-2), followed by a two-step excitation transfer to the right QD via $t_{d}$ (2-4), which finally decays via its own microwave gate (4-5).}
\label{fig:bspt}
\end{figure}

The many-body eigenstates of the uncoupled and undriven QD are the empty QD, $|0\rangle$, the single-electron doublet, $|\sigma\rangle=d^{\dagger}_{\sigma}|0\rangle$, and the doubly-occupied QD, $|2\rangle=d^{\dagger}_{\uparrow}d^{\dagger}_{\downarrow}|0\rangle$, with energies $0$, $\varepsilon_{d}$, $\varepsilon_{d}$ and $2\varepsilon_{d}$, respectively. For the proximitized QD, the BCS-like ground state becomes $|\tilde{0}\rangle=u|0\rangle+v|2\rangle$, the excited doublet, $|\sigma\rangle$ remains unchanged, and the highest excited state becomes $|\tilde{2}\rangle=u|2\rangle-v|0\rangle$, with energies $0$, $E_{d}$, $E_{d}$, and $2E_d$, respectively (cf. Fig.~\ref{fig:bspt}a).

In contrast to electric charge, the total DQD parity (odd/even number of quasiparticles) is conserved in the infinite-gap limit. Switching between even-parity states, $|\tilde{0}/\tilde{2}\rangle$, and odd-parity states, $|\sigma\rangle$, therefore takes place exclusively by inter-dot tunnelling with amplitude $t_d$. If the undriven system is prepared in its even-parity ground state, this will give rise to a finite Josephson current to second order in $t_d$, found from Eq.~\eqref{eq:td2curr} to be
\begin{align}
I=\frac{e|t_d|^{2}\Gamma^{2}}{(\varepsilon^{2}_{d}+\Gamma^{2})^{3/2}}\sin({\varphi_{sc}}).
\end{align}

In the driven case, a similar formula for the time-averaged current valid to second order in $t_{d}$ can be obtained when the system is driven with low amplitude, $A\ll E_{d}$, close to resonance, i.e. $\Omega\approx 2E_d$, as indicated in Fig.~\ref{fig:bspt}b. Since the mixing term~\eqref{eq:mix} is already proportional to driving amplitude, $A$, we shall neglect the time dependence of $E_{d}(t)$ in its denominator and in coherence factors $u$ and $v$, assuming that $A\ll\max(\epsilon_{d},\Gamma)$. This allows us to include the mixing term~\eqref{eq:mix} within a rotating wave approximation (RWA), which leads to the following equation of motion
\begin{equation}
i\frac{d}{dt}\chi_{\nu}(t)\approx \left[E_{d}\sigma^{z}_{\nu\nu'}+g e^{-i(\Omega t+\theta)\sigma^{z}_{\nu\nu}}\sigma^{x}_{\nu\nu'}\right]\chi_{\nu'}(t),
\end{equation}
where $g=A\Omega\Gamma/(2E_{d})^{2}$. This equation is readily solved by
\begin{align}
\chi_{\nu}(t)=e^{-i\sigma^{z}_{\nu\nu}(\Omega t+\theta)/2}
{\tilde U}^{-1}_{\nu\mu}\zeta_{\mu}(t),
\end{align}
with a secondary unitary transformation as
\begin{align}
{\tilde U}^{-1}_{\nu\nu'}=
\left(
\begin{array}{cc}
{\tilde u} & {\tilde v} \\
-{\tilde v} & {\tilde u} \\
\end{array}
\right)_{\!\!\nu\nu'},
\end{align}
defined in terms of
\begin{align}
{\tilde u}=\sqrt{(1+\delta/\tilde{E})/2},\hspace*{5mm}
{\tilde v}=\sqrt{(1-\delta/\tilde{E})/2}.
\end{align}
Here, $\delta=E_{d}-\Omega/2$ is the detuning, and the energy $\tilde{E}=\sqrt{\delta^{2}+g^{2}}$ captures the slow time evolution of the co-rotating Nambu spinor
\begin{align}
\zeta_{\mu}(t)=\zeta_{\mu}(0)e^{-i\tilde{E}t\sigma^{z}_{\mu\mu}},
\end{align}
with initial condition
$\zeta_{\mu}(0)=e^{i\sigma^{z}_{\nu\nu}\theta/2}{\tilde U}_{\mu\nu}\chi_{\nu}(0)$.

For concreteness, we assume that driving is turned on at time $t=0$, prior to which each proximitized QD is assumed to be thermalized in its ground state, $|\tilde{0}\rangle$. Using the relations,
\begin{align}
\chi_{\nu}(0)|\tilde{0}\rangle=\delta_{\nu,2}|\downarrow\rangle,\hspace*{5mm}
\chi^{\dagger}_{\nu}(0)|\tilde{0}\rangle=\delta_{\nu,1}|\uparrow\rangle,
\end{align}
the time-evolved states are found as
\begin{align}
\chi_{\nu}(t)|\tilde{0}\rangle=X_{\nu}(t)|\downarrow\rangle,\hspace*{5mm}
\chi^{\dagger}_{\nu}(t)|\tilde{0}\rangle&=i\tau^{y}_{\nu\nu'}X_{\nu'}(t)|\uparrow\rangle,
\end{align}
with
\begin{align}
X(t)=
\left(
\begin{array}{c}
e^{-i\Omega t/2}e^{-i\theta}i(g/\tilde{E})\sin(\tilde{E} t) \\
e^{i\Omega t/2}\left(\cos(\tilde{E} t)+i(\delta/\tilde{E})\sin(\tilde{E} t)\right) \\
             \end{array}
\right).
\end{align}
Reinstating the lead index $\alpha$ on $\theta$ and inserting this into Eqs.~\eqref{eq:Glg}, one finally arrives at the correlation functions
\begin{align}
G^{<}_{\alpha\eta,\alpha\eta'}(t,t')&=
i\left(b_{\alpha}(t),a_{\alpha}(t)\right)_{\eta}\left(b^{\ast}_{\alpha}(t'),a^{\ast}_{\alpha}(t')\right)_{\eta'}\\
G^{>}_{\alpha\eta,\alpha\eta'}(t,t')&=
-i\left(a^{\ast}_{\alpha}(t),-b^{\ast}_{\alpha}(t)\right)_{\eta}\left(a_{\alpha}(t'),-b_{\alpha}(t')\right)_{\eta'},\nonumber
\end{align}
where
\begin{align}
%
a_{\alpha}(t)&=e^{i\theta_{\alpha}/2}(-v,u)_{\nu}X_{\alpha\nu}(t),\nonumber\\
b_{\alpha}(t)&=e^{i\theta_{\alpha}/2}(u,v)_{\nu}X_{\alpha\nu}(t),
\end{align}
with the phase factor in front introduced merely for convenience in formulas below.
The time-dependent current in Eq.~\eqref{eq:td2curr} may now be expressed as
\begin{align}
I(t)&=4e|t_{d}|^{2}{\rm Re}\!\Big[a_{L}^{\ast}(t)b_{R}^{\ast}(t)
\int_{0}^{t}\!\!\!dt'\Big(a_{L}(t')b_{R}(t')\\
&\hspace*{5mm}+b_{L}(t')a_{R}(t')e^{-i\varphi_{sc}}\Big)-(L\leftrightarrow R, \varphi_{sc}\leftrightarrow -\varphi_{sc})\Big],\nonumber
\end{align}
involving products like
\begin{align}
a_{L}(t)b_{R}(t)&= 
i\big(v^{2}e^{-i\theta_d/2}-u^{2}e^{i\theta_d/2}\big)
(g/\tilde{E})\sin(\tilde{E} t)\nonumber\\
&\hspace*{10mm}\times\big(\cos(\tilde{E} t)+i(\delta/\tilde{E})\sin(\tilde{E} t)\big)\nonumber\\
&\hspace*{-14mm}+uv\Big[
e^{-i(\Omega t+\theta_{L}+\theta_{R})/2}(g/\tilde{E})^{2}\sin^{2}(\tilde{E} t)
\nonumber\\
&\hspace*{-5mm}
+e^{i(\Omega t+\theta_{L}+\theta_{R})/2}
\big(\cos(\tilde{E} t)+i(\delta/\tilde{E})\sin(\tilde{E} t)\big)^{2}
\Big],\nonumber
\end{align}
with $\theta_d=\theta_{L}-\theta_{R}$. The last two terms contain fast oscillating phase factors, $e^{\pm i\Omega t/2}$, and will be strongly suppressed by the subsequent integrations with respect to $t$. Notice that it is only these fast oscillating terms which contain information about the average phases of the drives, i.e. the information about the initial time, chosen here as $t=0$, from which the initial states are being time-evolved. Retaining only the first slowly oscillating term, which only carries information about the phase difference, $\theta_{d}$, these products reduce to
\begin{align}
a_{L}(t)b_{R}(t)&\approx -f(\theta_d)h(t),\nonumber\\
a_{R}(t)b_{L}(t)&\approx -f(-\theta_d)h(t),
\end{align}
with
\begin{align}
f(\theta_d)=&\frac{\varepsilon_{d}}{E_{d}}\cos(\theta_d/2)+i\sin(\theta_d/2),\nonumber\\
h(t)=&\frac{\delta g}{\tilde{E}^{2}}\sin^{2}(\tilde{E} t)-i\frac{g}{2\tilde{E}}\sin(2\tilde{E} t).
\end{align}
Finally, introducing $\kappa(\theta_d,\varphi_{sc})=f(\theta_d)e^{i\varphi_{sc}/2}=\kappa'+i\kappa''$, the current takes the following form
\begin{align}
I(t)
&\approx 8e|t_{d}|^{2}{\rm Re}\Big[\kappa^{\ast}(\theta_d,\varphi_{sc})h^{\ast}(t)\int_{0}^{t}\!\!\!dt'h(t')\Big]\kappa'(\theta_d,\varphi_{sc})\nonumber\\
&\hspace*{10mm}-(\theta_d\leftrightarrow -\theta_d, \varphi_{sc}\leftrightarrow -\varphi_{sc})\nonumber\\
&=2e|t_{d}|^{2}\frac{\delta g^{2}}{\tilde{E}^{4}}
\kappa'(\theta_d,\varphi_{sc})
\kappa''(\theta_d,\varphi_{sc})\\
&\hspace*{10mm}\times
\left[2\tilde{E}t\sin(2\tilde{E}t)-\sin^{2}(2\tilde{E}t)-4\sin^{4}(\tilde{E} t)\right].\nonumber
\end{align}
Whereas the first term oscillates around zero, the two last terms give rise to a well-defined long-time average
\begin{align}
\lim_{T\rightarrow\infty}\frac{1}{T}\int_{0}^{T}\!\!\!dt\,   \left(\sin^{2}(2\tilde{E}t)+4\sin^{4}(\tilde{E} t)\right)=2,
\end{align}
which results in the time-averaged current
\begin{align}\label{eq:ptcurr}
\langle I\rangle
&=-4e|t_{d}|^{2}\frac{\delta g^{2}}{\tilde{E}^{4}}
\kappa''(\theta_d,\varphi_{sc})\kappa'(\theta_d,\varphi_{sc})\nonumber\\
&\approx\frac{4e|t_{d}|^{2}A^{2}\Gamma^{2}}{E_{d}^{2}}
\frac{(\Omega-2E_{d})}{\left[(\Omega-2E_{d})^{2}+(A\Gamma/E_{d})^{2}\right]^{2}}\\
&\hspace*{5mm}\times
\left(\frac{\varepsilon_{d}^{2}}{E_{d}^{2}}\cos^{2}(\theta_d/2)+\sin^{2}(\theta_d/2)\right)\sin(\varphi_{sc}+\varphi_{0}),\nonumber
\end{align}
valid to leading order in $t_d$, close to resonance, $\Omega\approx 2E_{d}$, and with the phase shifted by
\begin{align}
\varphi_{0}=\arctan\left(\frac{2\varepsilon_{d}E_{d}\tan(\theta_{d}/2)}{\varepsilon_{d}^{2}-E_{d}^{2}\tan^{2}(\theta_{d}/2)}\right).
\end{align}
This current vanishes at resonance, $\Omega=2E_{d}$, and attains its maximum for $\Omega=2E_{d}\pm A\Gamma/(\sqrt{3}E_{d})$ with maximum current given by
\begin{align}
\langle I\rangle_{\rm max}&
\approx\frac{3\sqrt{3}e|t_{d}|^{2}E_{d}}{4A\Gamma}
\left(\frac{\varepsilon_{d}^{2}}{E_{d}^{2}}\cos^{2}(\theta_d/2)+\sin^{2}(\theta_d/2)\right)\nonumber\\
&\hspace*{15mm}\times
\sin(\varphi_{sc}+\varphi_{0}),
\end{align}
which is not strictly valid, since the maximum is attained where $\delta\sim g$, and counter rotating terms no longer are negligible.

Tuning the levels away from the Fermi level, i.e. for $|\varepsilon_{d}|\gg\Gamma$, we have $E_{d}\approx \varepsilon_{d}$ and the current becomes
\begin{align}
\langle I\rangle\approx&\,
\frac{4e|t_{d}|^{2}A^{2}\Gamma^{2}}{\varepsilon_{d}^{2}}
\frac{(\Omega-2|\varepsilon_{d}|)\sin(\varphi_{sc}+{\rm sgn}(\varepsilon_{d})\theta_d)}{\left[(\Omega-2|\varepsilon_{d}|)^{2}+(A\Gamma/\varepsilon_{d})^{2}\right]^{2}},
\end{align}
which is the Floquet $\theta_d$-junction, in which the phase shift of the sinusoidal current phase relation is set directly by the phase shift of the two driving fields together with the sign of the level energies set by $\varepsilon_{d}$. 

In the opposite limit, where the two levels are close to the Fermi levels of the two superconducting leads, i.e. $\varepsilon_{d}\ll\Gamma$, we have $E_{d}\approx\Gamma$ and arrive at
\begin{align}
\langle I\rangle&\approx4e|t_{d}|^{2}A^{2}
\frac{(\Omega-2\Gamma)\sin^{2}(\theta_d/2)}{\left[(\Omega-2\Gamma)^{2}+A^{2}\right]^{2}}\sin(\varphi_{sc}),
\end{align}
which corresponds to a $0$-junction right above resonance where $\Omega>2\Gamma$, and a $\pi$-junction right below resonance for $\Omega<2\Gamma$. In this limit, the phase shift of the two drives, $\theta_d$, serves only to modulate the amplitude, attaining maximum average current when the drives are shifted by $\theta_{d}=\pi$, and blocking it altogether for $\theta_{d}=0$.

This average current was calculated under the assumption of an even number of electrons occupying each of the two levels, with the specific initial condition that the system is in its lowest energy state at time zero. In a real system, however, quasiparticle poisoning, and relaxation will cause occasional switching of the parity of each of the two levels. With typical parity flip times of the order of 20–200 $\mu$s~\cite{Janvier2015Sep, Hays2018Jul, Hays2020Nov, Hays2021Jul}, a resonant drive frequency, $\Omega\sim 2E_{d}$, of the order of 10 GHz, say, will take the system through some 10$^6$ cycles before the parity is flipped, implying that the average current determined here remains meaningful as long as $\tilde{E}\gg 10^{-6}\Omega$, i.e. $\sqrt{A\Gamma}\gg 10$ MHz. The full problem thus entails the stochastic element of random parity switching between even, and odd parity sectors of the Hilbert space. This poses an interesting problem in itself, but shall not be pursued any further in this work. Instead, we shall analyze the steady state Dyson equation~\eqref{eq:GLDyson}, in which the parity is relaxed in the infinite gap limit by a weak coupling to a normal metallic reservoir. For a finite gap, the Floquet sidebands of the continuum provide the same effect and the normal metallic reservoir is no longer needed.

Notice that the full lesser component of the Dyson equation has a second contribution~\cite{Haug2008}, $(1+G^R\Sigma^R)G_0^<(1+\Sigma^{A}G^{A})$, referring to the initial lesser function, and that this term has been omitted altogether in equation~\eqref{eq:GLDyson}. This omission rests on the tacit assumption, that $\Sigma^{<}$ contains relaxation mechanisms, which will wash out the initial conditions, i.e. that $\Sigma^{<}\gg G^{R,-1}_{0}G_0^<G^{A,-1}_{0}=(G^{R,-1}_{0}-G^{A,-1}_{0})f_{0}$, where $f_0$ denotes an initial distribution function. In the present tunnelling problem, $\Sigma^<$ refers to quasiparticle tunnelling to and from the superconducting leads and to the weak tunnelling of electrons directly between the dots and a normal metal reservoir. The former contribution vanishes altogether in the infinite-gap limit, and the steady state Dyson equation~\eqref{eq:GLDyson} as well as the Floquet Keldysh transformation~\eqref{eq:FLtransf}, is therefore justified in the infinite gap limit by the normal metal tunneling rate, $\Gamma_m$, which is large enough to dominate the finite $\eta=(G^{R,-1}_{0}-G^{A,-1}_{0})/(2i)$ used in our numerical implementation of the bare Green functions of the leads, yet small enough not to affect the result.

\section{Numerical results}~\label{sec:numerics}

In this section we present numerical results obtained with the Floquet Keldysh Green functions introduced in Sec.~\ref{sec:Greens}. We shall focus entirely on the time-averaged quantities, which may be found as the zero'th Floquet components, and we shall narrow down the rather large parameter space to illustrate some of the most interesting time-averaged current phase relations realized by this driven junction.

In practice, the inversion of~\eqref{eq:GRinv} is carried out by truncating to the $n_{\rm max}$ lowest Floquet modes, i.e. working with square matrices of dimension $4(1+2n_{\rm max})$. For all numerical results presented below, we ensure that $n_{\rm max}$ is large enough that increasing it further does not affect the results. Furthermore, we use a finite broadening in the lead Green functions, replacing $0_{+}$ by $\eta=10^{-4}$ in Eq.~\eqref{eq:gR}, which, like all energy and frequency ($\hbar\equiv 1$) parameters used below (except for the infinite-gap limit), is specified in units of $\Delta$. In order to facilitate the numerical integration over the sharp sub-gap states in the infinite gap limit, both levels are weakly coupled to a normal metallic lead with chemical potential aligned with the two superconducting leads, $\mu_{m} = 0$. This gives rise to a finite imaginary part, $\Gamma_{m}$, of the $d$-electron self-energies~\eqref{eq:selfen}, which is chosen to be smaller than any other scale in the problem, yet resolved by the discretized numerical integrations. In practice, this corresponds to a finite parity relaxation time, which is longer than any other timescale in the problem. As discussed in the previous section, this also constitutes the formal justification of the steady state Dyson equation~\eqref{eq:GLDyson}. For a finite gap, the continuum of the superconducting leads provides the necessary broadening for the numerical calculations, and the normal metallic lead is not needed.

\begin{figure}
\centering
\includegraphics[width=\columnwidth]{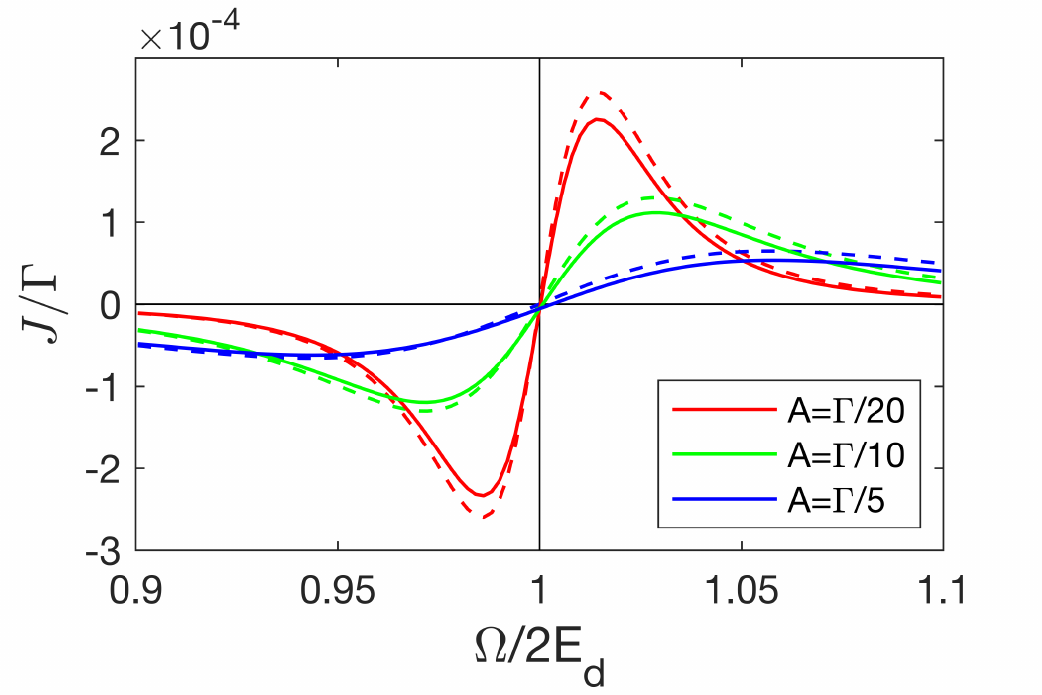}
\caption{{\it (Dashed lines)} Plots of the weak coupling anomalous Josephson ($\varphi_{sc}=0$) current in Eq.~\eqref{eq:ptcurr} in the infinite-gap limit as a function of the driving frequency, showing the maxima on each side of the node right at resonance, $\Omega=2E_{d}$. Parameters are chosen such that $t_d =\Gamma/100$, $\varepsilon_d = \Gamma/10$ and $\theta_{d}=\pi/2$, together with three different driving amplitudes (see inset). {\it (Full lines)} Numerical calculation of the current using Eq.~\eqref{eq:current_keldysh} with same parameters as above and with $\Delta = 2\times 10^4$ , $\eta = 2\times 10^{-4}$ and an additional broadening of the QD states corresponding to a normal metal tunnelling rate $\Gamma_m = \Gamma/500$.}
\label{fig:ptcompare}
\end{figure}

\subsection{Infinite, and large-gap results}

\begin{figure}
\centering
\includegraphics[width=\columnwidth]{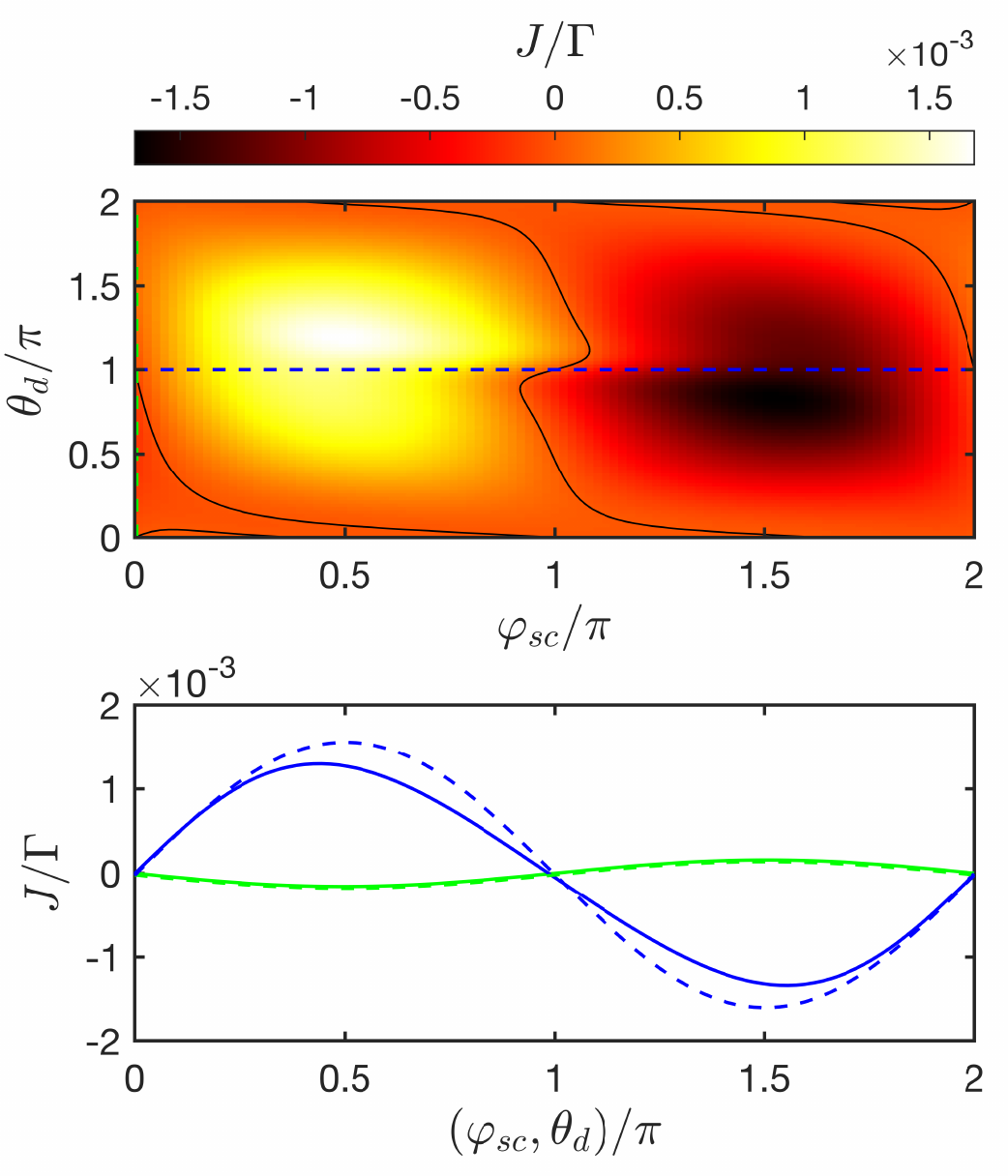}
\caption{{\it (Upper panel)} Density plot of the weak-coupling current vs. superconductor phase difference and phase shift of the two drives obtained by numerical evaluation of Eq.~\eqref{eq:current_keldysh}. {\it (Lower panel)} Solid lines correspond to cuts along the dashed blue ($\theta_{d}=\pi$), and green ($\varphi_{sc}=0$) lines indicated in the upper panel, together with the corresponding analytical infinite-gap weak coupling current from Eq.~\eqref{eq:ptcurr} limit (dashed). In both panels, parameters are $t_d=0.02\Gamma, A=\varepsilon_d=0.1\Gamma$ and $\Omega=2\Gamma$. In the numerical evaluation the infinite gap was replaced by $\Delta = 10^4$, while $\eta = 10^{-4}$ and $\Gamma_m = \Gamma/300$.}
\label{fig:phitheta}
\end{figure}

In order to connect to the results of the previous section, we first consider the infinite-gap limit, in which all current is carried by Cooper pairs, at weak tunnel coupling and close to resonance. The resulting current, $J=J_{L}^{0}-J_{R}^{0}$ (see Eq.~\eqref{eq:current_keldysh}), is shown in Fig.~\ref{fig:ptcompare} for different frequencies around the resonance. It is seen to match the perturbative results very well. We plot in Fig.~\ref{fig:phitheta} the dependence of the current near the resonance on the two phases, $\varphi_{sc}$ and $\theta_d$.We show this together with two cuts illustrating a good match with the result obtained in Eq.~\eqref{eq:ptcurr}.

\begin{figure}
\centering
\includegraphics[width=\columnwidth]{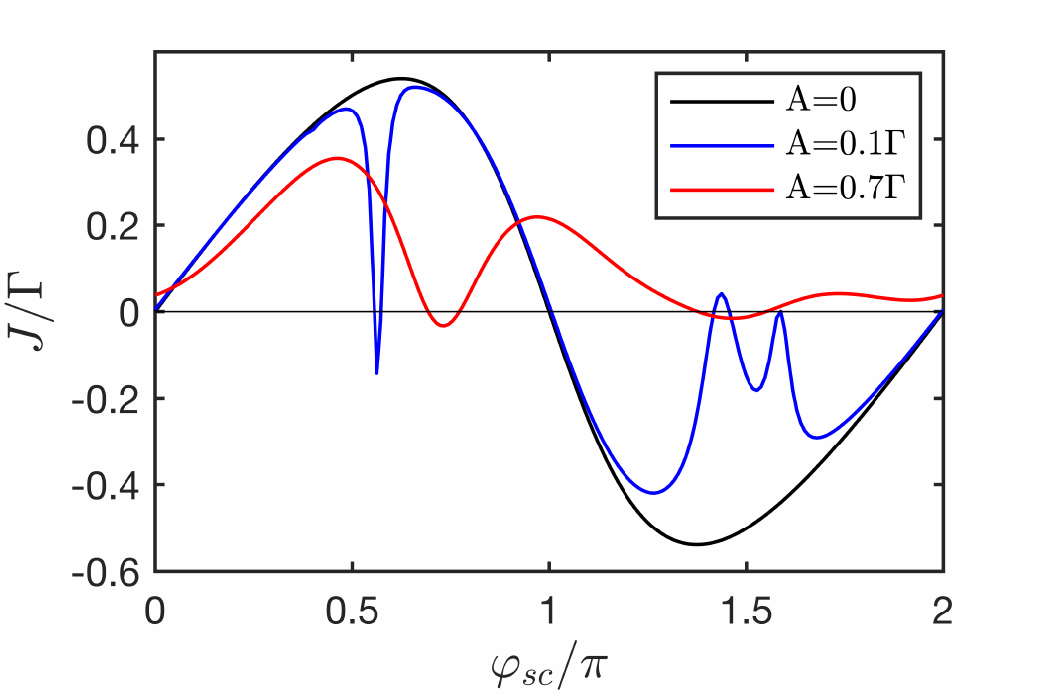}
\caption{Current phase relation (CPR) in the infinite-gap limit ($\Delta=2\times10^{4}\Gamma$) with, and without drives of amplitude $A$, frequency $\Omega = 2.2\Gamma$ and phase shift $\theta_d=\pi/2$. Both levels have energy $\varepsilon_d = 0.8\Gamma$ with a weak normal metal tunnelling rate $\Gamma_m = \Gamma/300$, and are tunnel coupled by $t_d = 2\Gamma$.} 
\label{fig:posj}
\end{figure}

Increasing the amplitude of the drive and fixing the driving phase shift at $\theta_{d}=\pi/2$, gives rise to highly non-trivial CPRs, of which a few examples are shown in Fig.~\ref{fig:posj}. For a small driving amplitude the CPR is modified by narrow dips of the current, similar to what is observed in superconducting junctions with only a single drive~\cite{Bergeret2010,Bergeret2011,Cuevas1996,Cuevas2002,Martin-Rodero1999}, the main difference being that in this case the dips are not symmetric around $\varphi_{sc}=\pi$ and do not reach zero, similar to the phase-shifted results in Ref.~\onlinecite{Venitucci2018}. For higher driving amplitudes ($A=0.8\Gamma$ shown) the current is reduced, as for the junctions with only a single drive, but now the CPR is severely modified with no special significance of neither $\varphi_{sc}=0$ nor  $\varphi_{sc}=\pi$, both exhibiting finite supercurrent.
For comparison, in Appendix~\ref{app} we calculate the current using the same parameters as for the blue curve ($A=0.1\Gamma$) in Fig.~\ref{fig:posj}, but now using Floquet states to determine the time evolution of the non-driven ground state. This is done in the infinite-gap limit and with no coupling to a normal metal ($\Gamma_{m}=0$). The long-time average of the resulting current shows good correspondence with the steady-state current in Fig.~\ref{fig:posj}. Furthermore, interactions are straightforwardly included in this approach, and are shown to remove the sharp dips in the current if $U$ becomes of the order of the driving frequency $\Omega$.
\begin{figure*}[t]
\centering
\includegraphics[width=2\columnwidth]{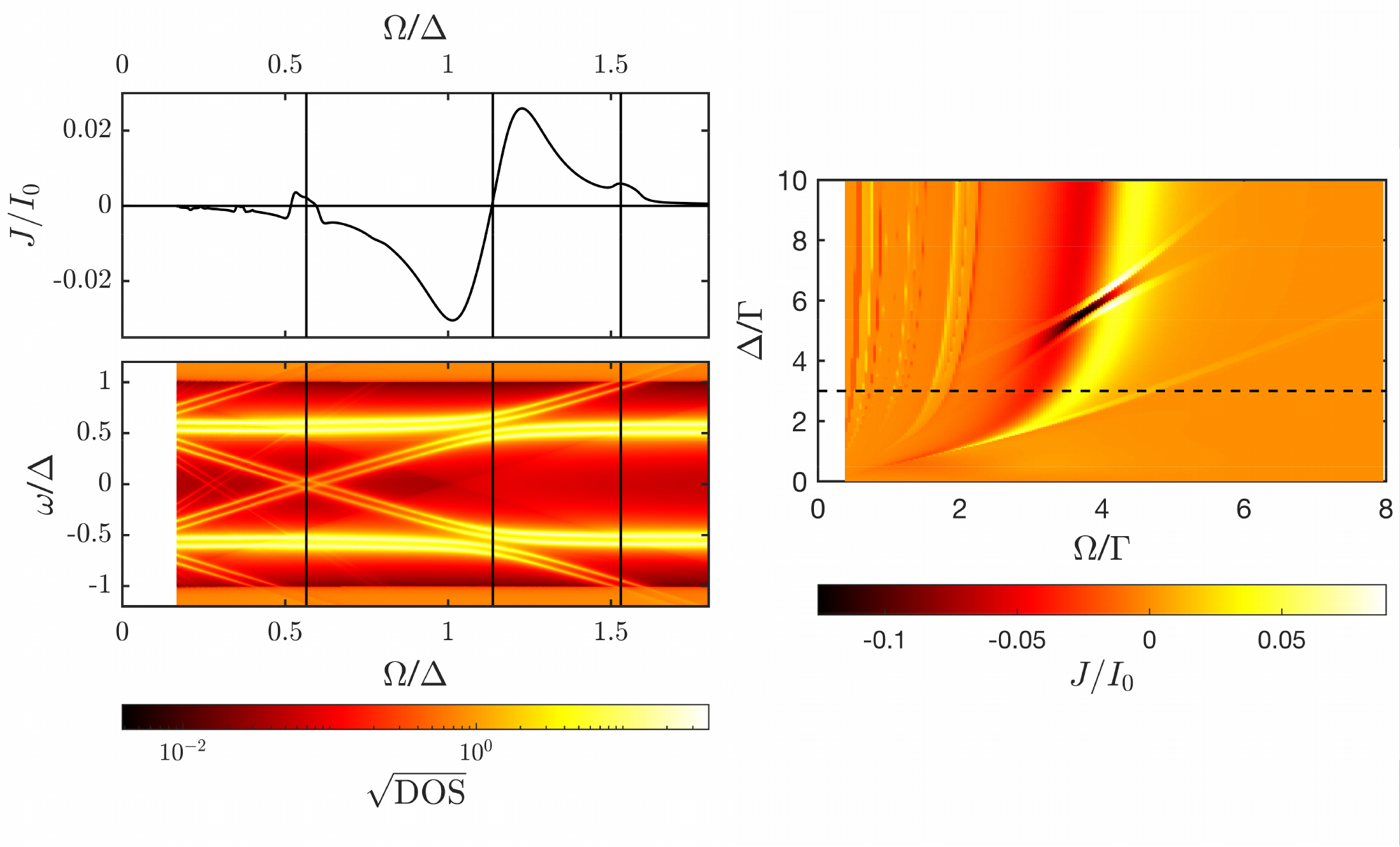}
\caption{(Left) Upper panel: The time-averaged current (in units of $I_0=2e\Delta$) at zero superconductor phase difference ($\varphi_{sc}=0$) as a function of drive frequency. Lower panels: the corresponding time-averaged density of states on the two resonant levels. Both panels are evaluated with $\Gamma =\Delta/3, t_d = 2\Delta/3, A = 4\Delta/15, \varepsilon_d = \Delta/15$ and $\theta_d=\pi/2$.
(Right) Variation with $\Omega$ and $\Delta$ of the pumped current with $t_d = 2\Gamma, A = 0.8\Gamma, \varepsilon_d = 0.2\Gamma,\varphi_{sc}=0,\theta_d=\pi/2$ and $n_{max}=7$. For high $\Delta$ resonances are observed between different ABS, while for low $\Delta$ the current is featureless, signaling an adiabatic origin of the current. The line separating these two regions corresponds to the resonance between the ABS and the continuum. A second line with twice the slope is also present, corresponding to the first harmonic of that resonance.
The dashed line at $\Delta = 3\Gamma$ is the cut shown in Fig.~\ref{fig:resonance}.}
\label{fig:resonance}
\end{figure*}
\twocolumngrid
The systematic behavior relies on many parameters, but the interaction seems to merely change the resonance condition for the drive.

\subsection{Finite gap results}

Turning to the case of a finite BCS gap, $\Delta$, we first fix the superconductor phase difference to zero ($\varphi_{sc}=0$) and calculate the frequency dependence of the time-averaged current. This is shown in the upper left panel of Fig.~\ref{fig:resonance}. The lower left panel shows the corresponding time-averaged density of states on the proximitized dots, exhibiting pronounced peaks at two slightly different ABS energies, together with their weaker first, and even weaker second Floquet sidebands. The small peak in the current at the highest frequency corresponds to a resonance between the first side band of the two ABS and the BCS quasiparticle continuum, at around $\Omega = \Delta + E_{ABS}\simeq 1.5\Delta$. At lower frequencies the current attains its largest magnitude slightly off-resonance, and a node right at resonance, $\Omega \simeq 2E_{ABS} \simeq  1.2\Delta$. Here a positive ABS energy matches the first sideband of a negative ABS energy, or vice-versa, as illustrated in Fig.~\ref{fig:bspt} for weak $t_{d}$. For lower frequencies, crossings of ABS sidebands with each other or with the continuum are again reflected in the current. Apart from this additional structure arising from the finite gap or from a substantial driving amplitude, the overall frequency dependence of the current clearly resembles the resonant structure found in the infinite-gap limit in Fig.~\ref{fig:ptcompare}. For the rest of the paper, we fix the drive frequency to be slightly deviated from the main resonance at $\Omega \simeq 2E_{ABS}$ so as to focus our attention to this anomalous supercurrent signal.

To further investigate the effect of the continuum in the anomalous current we show in the right panel of Fig.~\ref{fig:resonance} how the current varies with $\Delta$ and $\Omega$. Coming from high $\Delta$, the resonant min-zero-max structure observed in Fig.~\ref{fig:ptcompare} and in the upper left panel of Fig.~\ref{fig:resonance} persists down to a ration of  approximately $\Delta/\Gamma \simeq 2-3 $, with only a slight shift in the resonance frequency. For low $\Delta$ the current is very small and frequency independent, consistent with a weak adiabatic pumping of normal current, similar to the case with normal leads~\cite{Riwar2010}. The line separating the two regions corresponds to the condition for resonance between the continuum and the first Floquet sideband to the negative energy ABS, i.e. $\Omega = \Delta + E_{ABS}$, where the energy of the ABS itself also depends on the superconducting gap~\cite{Bauer2007, Bergeret2010, Haller2014}. A second line with twice the slope is also observed: it corresponds to a resonance between the second Floquet sideband and the quasiparticle continuum, beyond which the resonances carrying supercurrent are not modified. Interestingly, this second sideband is observed to anticross with the first sideband of the positive energy ABS, which gives rise to a large enhancement of the negative resonant current peak at frequency just below $\Omega=2E_{ABS}$. Since this anticrossing involves sidebands crossing with the continuum, this enhancement of the current is most likely due to a dissipative quasiparticle current. On the other hand, the nearly vertical features in this figure, including the most pronounced min-zero-max resonance, correspond to a current of Cooper pairs, which are being pumped across the junction by phase shifted resonances between sub-gap states and their Floquet sidebands like the one indicated in Fig.~\ref{fig:bspt}.

\subsection{Modified and rectifying current phase relations}

As established in the Appendix~\ref{app} for single-state time evolution in the infinite-gap limit, the symmetries of the Floquet Hamiltonian guarantee the following symmetries of the time-averaged current: 
\begin{align}
J(\varphi_{sc},\theta_d,\varepsilon_d)
&=- J(-\varphi_{sc},\theta_d,-\varepsilon_d),\label{eq:1s}\\
&=-J(-\varphi_{sc},-\theta_d,\varepsilon_d).\label{eq:2s}
\end{align}
As we shall see below, these symmetries are still obeyed in the steady state Green function calculations with a finite BCS gap.

\begin{figure}
\centering
\begin{subfigure}{}
    \includegraphics[width=\columnwidth]{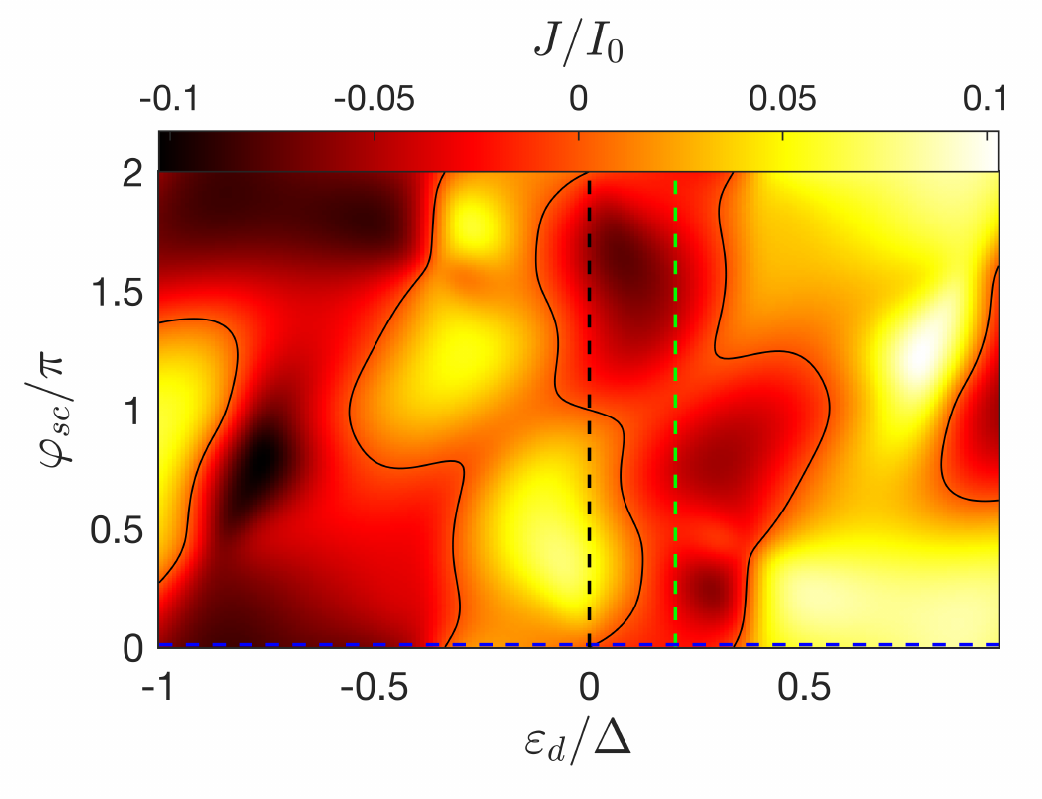}
    \label{fig:first}
\end{subfigure}
\vspace{-1.5cm}.
\begin{subfigure}{}
    \includegraphics[width=\columnwidth]{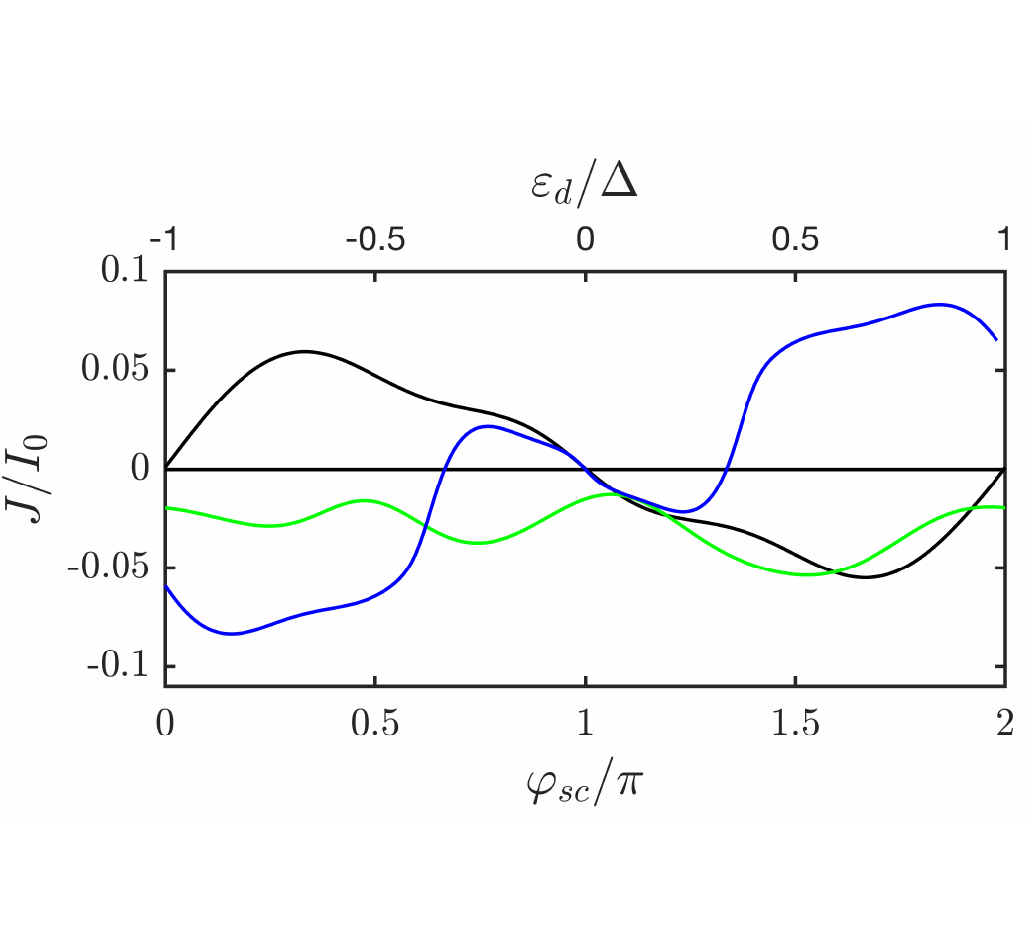}
    \label{fig:third}
\end{subfigure}
\vspace{-1cm}.
\caption{Variation with $\varepsilon_d$ and $\varphi_{sc}$ of the pumped current with parameters $2\Gamma = t_d = 0.7\Delta,A=0.8\Delta, \Omega = 0.9\Delta,\theta_d=\pi/2$ and $n_{max}=7$ . The current-phase relation is strongly modified by varying $\varepsilon_d$ and the antisymmetry for inversion of $\varphi_{sc}$ and $\varepsilon_d$.  As in Fig 6, another case is shown in green. here the current-phase relation does not cross zero current. The blue curve shows, for $\varphi_{sc}=0$, the variation of the pumped current with $\varepsilon_d/\Delta$.}
\label{fig:eda}
\end{figure}

The first symmetry relation,~Eq.\eqref{eq:1s}, is apparent in the numerical Green function result for the time-averaged current shown in Fig.\ref{fig:eda}. For $\varepsilon_d=0$, the vertical black dashed cut illustrates the usual antisymmetry around $\varphi_{sc}=\pi$. This symmetry breaks down for $\varepsilon_{d}\neq 0$ and, as indicated by the vertical green dashed cut, may even lead to a unidirectional supercurrent, corresponding to {\it complete rectification}. The horizontal black dashed cut, on the other hand, illustrates the antisymmetry of the current under inversion of $\varepsilon_d$ for $\varphi_{sc}=0$.

\begin{figure*}[t]
\centering
\includegraphics[width=2 \columnwidth]{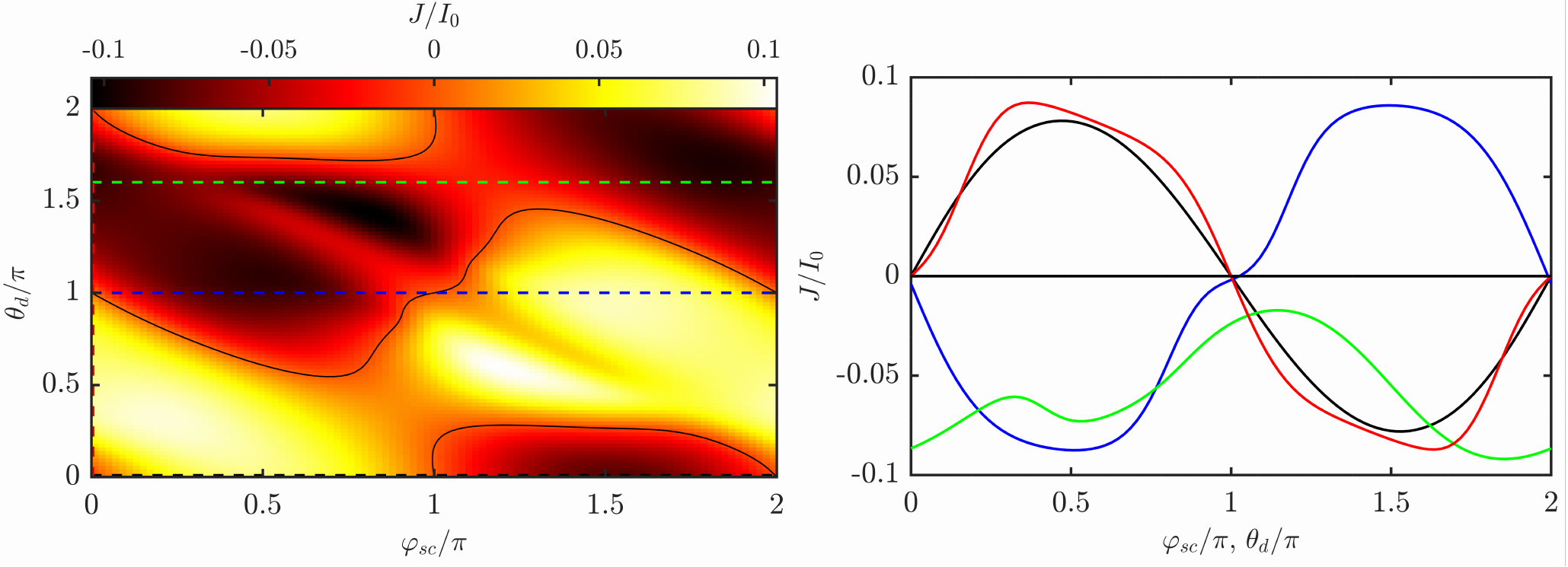}
\caption{Variation with $\varphi_{sc}$ and $\theta_d$ of the pumped current with $2\Gamma = t_d = 0.7\Delta, A = 0.8\Delta, \varepsilon_d = 0.8\Delta, \Omega=0.9\Delta$ and  $n_{max}=11$. The current-phase relation is strongly modified and it shows antisymmetry for inversion of both phases. The current vanishes at the black solid lines. We show four cuts (marked by dashed lines) where interesting features are observed. In black, with $\theta_d=0$, the current is that of a $\pi-$junction while for $\theta_d=\pi$, shown in blue, the junction becomes a $0-$junction again. Another interesting feature is shown in green, with $\theta_d = 1.6\pi$, where for certain parameters one can obtain a current-phase relation that does not vanish for any value of the superconducting phase.}
\label{fig:phiphi}
\end{figure*}

In Fig.~\ref{fig:phiphi} we illustrate the second symmetry relation~\eqref{eq:2s} of the current under inversion of both phases, $\varphi_{sc}$ and $\theta_d$. The vertical red dashed cut shows the anomalous relation between time-averaged current and phase difference on the drives, $\theta_{d}$, at a superconductor phase-difference fixed at $\varphi_{sc}=0$, attaining its maximum near, but not right at $\theta_{d}=\pi/2$. The three horizontal (black, blue and green) dashed cuts illustrate the the strongly modified current phase relations between the time-averaged current and the superconductor phase-difference. Switching from $\theta_d = 0$ to $\theta_d = \pi$, the driven Josephson junction is seen to switch current phase relation from a $0$-, to a $\pi$-junction, as seen in the black, and the blue curves, respectively, up to a slight anharmonicity in both.

Once again, the green cut realizes a rectified time-averaged current. Since the BCS gap is finite, there is no guarantee that this completely rectified pump current is exclusively a current of Cooper pairs. Nevertheless, as we show in Fig.~\ref{fig:AppFig3} in Appendix~\ref{app}, a completely rectified current can also be obtained in the infinite-gap limit where all current must be carried by Cooper pairs, indicating that there is no fundamental obstacle to attaining a unidirectional time-averaged supercurrent for all $\varphi_{sc}$.

\twocolumngrid
\section{Discussion}~\label{sec:discuss}

As demonstrated above, bridging two superconductors by a double quantum dot with phase shifted microwave tones on their respective gate voltages, as depicted in Fig.~\ref{fig:setup_2qd}, comprises an effective Josephson junction with a highly nontrivial CPR. More specifically, the driving induces an alternating tunnelling current, which may exhibit a well-defined non-zero long-time average, and it is this average current, which exhibits an anomalous and often highly anharmonic relation to the superconductor phase difference. In light of the recent interest in Josephson diodes~\cite{Ando2020Aug,Wu2022Apr,Souto2022May}, it is worth stressing that this driven junction offers complete rectification of the time-averaged supercurrent.

The supercurrent response to the driving relies on non-adiabatic resonant photon assisted tunnelling. This was established in the infinite-gap limit by means of perturbation theory and by time evolution of the non-driven ground state using Floquet theory (cf. Appendix~\ref{app}). For a finite BCS gap, a steady state time-averaged current was calculated by means of Floquet Keldysh Green functions, for which a weak tunnel coupling of each dot to a normal metal was included to eliminate the transient response and allow for parity relaxation. Whereas the general finite-gap current may include some fraction of BCS quasiparticles, the main resonant pump current arising when the drive frequency is slightly off resonance with the energy difference between the two even-parity sub-gap ABS was argued to be carried mainly by Cooper pairs. 

For clarity, we have restricted our analysis to a symmetrically coupled device, where only the microwave phase shift breaks the $L/R-$inversion symmetry. Even in this case, the mean current exhibits a highly non-trivial behavior on the remaining parameters, such as the common mean gate voltage, the inter-dot tunnel coupling together with amplitude and frequency of the microwave tones. The plots chosen to illustrate the salient features for this work therefore by no means exhaust the many possible behaviors of this driven DQD junction. With two different tunnel couplings to the two superconductors, the induced ABS will have different energies and consequently the resonance frequencies on the two quantum dots will be different. The mechanism underlying the resonant current response will, however, remain viable if the two drive frequencies may be adjusted independently. 

As demonstrated for the infinite-gap limit in Appendix~\ref{app}, local Coulomb interactions, reflecting the finite charging energies of the quantum dots were shown to alter the resonance conditions and thereby affect the time-averaged current. Nevertheless, the anomalous Josephson effect (and the rectification) persisted, and was found to exhibit a $0-\pi$ transition in $\theta_{d}$, as the interaction strength increased past a critical value. In real systems, this analysis pertains to the weak coupling regime, $U\ll\Delta$, whereas the opposite regime of $U\gg\Delta$ leads to the formation of YSR states. In this case, quasiparticles from the BCS continua in the leads form singlet bonds with spinful (odd-occupied) quantum dots, and this incomplete proximity effect must be expected to lead to substantial pumping of quasiparticle current. In the limit of $U\gg\Gamma,t_{d}$, however, charge fluctuations on the dot will be strongly impeded and the driving only effective at higher frequencies and amplitudes. It should be interesting to explore this regime further. Such work would also allow extending the results of Ref.~\onlinecite{Hermansen2022Feb} to a setup for two-tone spectroscopy~\cite{Metzger2021Jan,Bargerbos2022Feb}.

As discussed briefly at the end of Section~\ref{infgap}, the time-averaged currents addressed in this work must be expected to depend strongly on the parity flip dynamics arising from quasiparticle poisoning in a given device. Here, we have restricted our attention to the NESS Green function approach or to parity conserving time evolution of a single state, which in spite of their obvious differences all agree on the salient features of the driven DQD junction. For future work, it would be instructive to add stochastic parity dynamics to the Floquet time evolution carried out in Appendix~\ref{app} and make a comparison with the NESS Green function results. 

The microwave enabled DQD Josephson junction studied here offers a highly tunable superconducting circuit element, which clearly links the phase-shifted AC input to a traversing supercurrent. Here we have only addressed the relations between time-averaged currents and superconducting, as well as microwave phase differences. To assess the possible value of such a circuit element, future work should address the time-dependent higher harmonics of the induced current as well as its possible implementation in a superconducting circuit.

In the course of finishing this work, we noticed the appearance of new work by A. Soori~\cite{Soori2022Jun}, who also points to the Josephson diode aspect present in the driven two-site SNS junction explored also in Ref.~\onlinecite{Soori2020Jun}.

{\it Acknowledgments.} The Center for Quantum Devices (Project No. DNRF101) and the Center for Nanostructured Graphene
(Project No. DNRF103) 
are funded by the Danish National Research Foundation. We acknowledge fruitful discussions with G. Steffensen, K. Flensberg and M. Geier.

\appendix
\section{Floquet analysis of the interacting infinite-gap limit}
\label{app}

The infinite-gap limit offers relatively easy access to the symmetries of the problem, which are also revealed by the steady-state numerical calculations presented in the main text. In this appendix, we employ Floquet theory to provide a brief supplementary analysis of this more tractable limit, in which Local Coulomb interactions on the quantum dots can readily be included. Furthermore, since no quasiparticle excitations are involved in the infinite-gap limit, all currents calculated below are carried exclusively by Cooper pairs. We choose to consider only the even-parity sector, but a similar analysis is straightforwardly made for the odd-parity sector.

In the even-parity sector, the Hilbert space is spanned by the basis,  $\{|00\rangle,|20\rangle,|02\rangle,|22\rangle, |\uparrow\downarrow\rangle,|\downarrow\uparrow\rangle\}$, where the left(right) index indicates the many-body states of the left(right) dot. In this basis the first quantized Hamiltonian reads
\begin{align}
\hat{H}_{e,\infty}=\left(
\begin{array}{cccccc}
 0 & -\Gamma & -\Gamma & 0 & 0 & 0  \\
-\Gamma & 2\varepsilon_d+U & 0 & -\Gamma & z & -z  \\
-\Gamma & 0 & 2\varepsilon_d+U  & -\Gamma & z^\ast & -z^\ast  \\
0 & -\Gamma & -\Gamma & 4\varepsilon_d+2U & 0 & 0  \\
0 & z^\ast & z & 0 & 2\varepsilon_d & 0  \\
0 & -z~\ast & -z & 0 & 0 & 2\varepsilon_d
\end{array}\right),
\end{align}
with tunnelling matrix elements $z=t_{d}e^{i\varphi_{sc}/2}$, and with the local intra-dot Coulomb interaction, $U$, now included. From this, one may construct the even-parity Floquet Hamiltonian, $H_{e}^{F}$ corresponding to the harmonic driving term, $A\cos(\Omega t)$, from the matrix elements~\cite{Sambe1973, Shirley1965, Eckardt2015}
\begin{align}
\hat{H}^F_{e,mn} =&\,\left(\hat{H}_{e,\infty}-n\Omega\hat{I}\right)\delta_{mn}
+\hat{V}\delta_{n-m,1}
+\hat{V}^{\dagger}\delta_{m-n,1},
\end{align}
where $\hat{I}$ denotes the $6\times 6$ unit matrix, and $\hat{V}$ is defined as the $6\times 6$ matrix with diagonal elements
\begin{align}
A\left\{0,e^{i\theta_{L}},e^{i\theta_{R}}, e^{i\theta_{L}}+e^{i\theta_{R}},\frac{e^{i\theta_{L}}+e^{i\theta_{R}}}{2},\frac{e^{i\theta_{L}}+e^{i\theta_{R}}}{2}\right\},
\end{align}
and zeros elsewhere. Truncating this infinite dimensional matrix and solving the $6(2n_{\rm max}+1)$ dimensional eigenvalue problem
\begin{align}
\sum_{n=-n_{max}}^{n_{max}}\hat{H}^F_{e,mn}|u_{\nu}^{n}\rangle=\epsilon_{\nu}|u_{\nu}^{m}\rangle,
\end{align}
the time-dependent Schr\"{o}dinger equation is solved by the 6 Floquet states,
\begin{align}
|\psi_{\nu}(t)\rangle=e^{-i\epsilon_{\nu}t}\sum_{n=-n_{max}}^{n_{max}}e^{-in\Omega t}|u_{\nu}^{n}\rangle,
\end{align}
corresponding to the 6 quasienergies in the first Floquet Brillouin zone, $-\Omega/2<\epsilon_{\nu}<\Omega/2$, for $\nu=1,2,\ldots,6$. Expressing these 6 eigenstates in the original 6-dimensional even-parity basis, $|u_{\nu}^{n}\rangle=\sum_{i}u_{\nu}^{n}(i)|i\rangle$, a given initial state may now be expressed as
\begin{align}
|\Psi(0)\rangle =\sum_{\nu,i=1}^{6}c_{\nu}\!\!\!
\sum_{n=-n_{\rm max}}^{n_{\rm max}}\!\!\!u_{\nu}^{n}(i)|i\rangle,
\end{align}
from where the coefficients $c_{\nu}$ are found by inverting the square ($n\nu$) matrices  $u_{\nu}^{n}(i)$. Finally, the solution for the  full time evolution of the state can be expressed as
\begin{align}
|\Psi(t)\rangle=\sum_{\nu,i=1}^{6}c_{\nu}e^{-i\epsilon_{\nu}t}\!\!\!\sum_{n=-n_{max}}^{n_{max}}\!\!\!e^{-in\Omega t}u_{\nu}^{n}(i)|i\rangle.
\end{align}

\subsection{Time-averaged current}

From this time-evolved state, the time-dependent expectation value of the current operator, $\hat{I}=2e\partial_{\varphi_{sc}}\hat{H}_{e,\infty}$, is determined as
\begin{align}
I(t)&=\langle\Psi(t)|\hat{I}|\Psi(t)\rangle\\
&=\!\!\!\!\sum_{\mu\nu,ij,mn}\!\!
c_{\mu}^{\ast}c_{\nu}
\left(u_{\mu}^{m}(j)\right)^{\ast}\!u_{\nu}^{n}(i)
e^{i(\epsilon_{\mu}-\epsilon_{\nu}+(m-n)\Omega)t}\langle j|\hat{I}|i\rangle,\nonumber
\end{align}
which leads to the long-time average
\begin{align}
J=\langle I\rangle&=
\lim_{T\rightarrow\infty}\frac{1}{T}\int_{0}^{T}\!\!\!dt\,
\langle\Psi(t)|\hat{I}|\Psi(t)\rangle\label{eq:flocurr}\\
&=\!\!\sum_{\nu,n,ij}\!
|c_{\nu}|^{2}\left(u_{\nu}^{n}(j)\right)^{\ast}\!u_{\nu}^{n}(i)\langle j|\hat{I}|i\rangle.\nonumber
\end{align}

\begin{figure}
\centering
\includegraphics[width=\columnwidth]{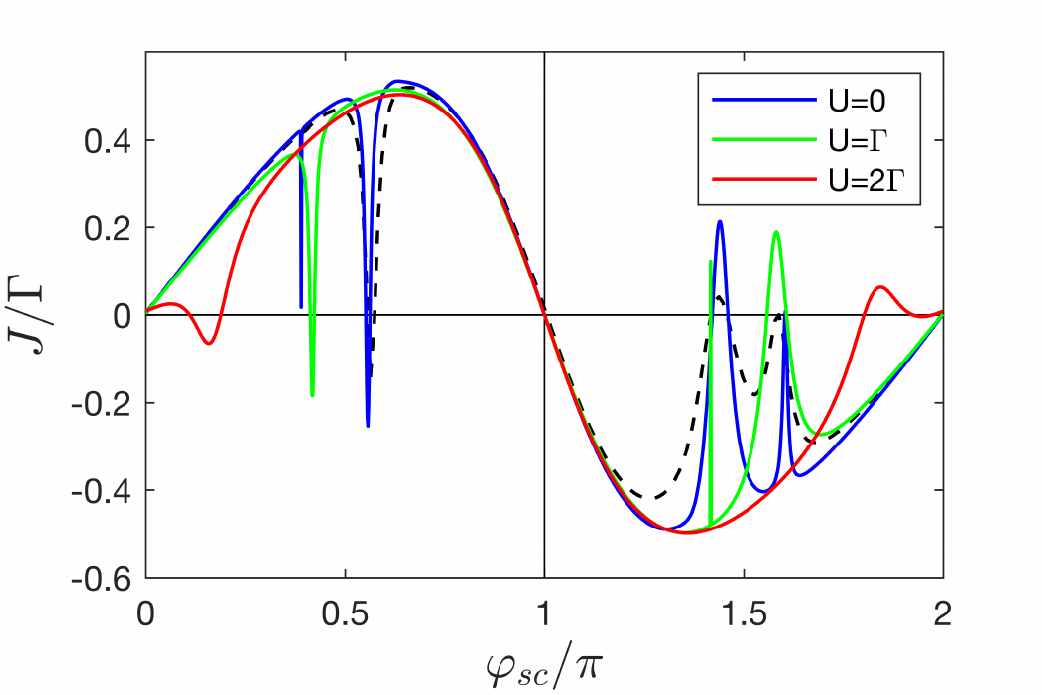}
\caption{Interaction dependence of the current-phase relations for parameters as in Fig.~\ref{fig:posj} ($\Omega = 2.2\Gamma$, $\theta_d=\pi/2$, $\varepsilon_d = 0.8\Gamma$, $t_d = 2\Gamma$), with $A=0.1\Gamma$. Full lines correspond to the result obtained using the methods in this appendix. Blue (green, red) lines correspond to $U=0$ $(U=\Gamma, U=2\Gamma$). Dashed line is the same result as Fig.5, also for $A=0.1\Gamma$.}
\label{fig:AppFig1}
\end{figure}

Using the same parameters as in Fig.~\ref{fig:posj} and choosing the ground state of the undriven system as the initial state, one may now calculate the matrices, $u_{\nu}^{n}(i)$, together with the corresponding coefficients, $c_{\nu}$, and evaluate the time-averaged current using formula~\eqref{eq:flocurr}. The result is shown in Fig.~\ref{fig:AppFig1}, with full (dashed) lines corresponding to the driven (undriven) system and the blue (green, red) lines corresponding to $U=0$ $(U=\Gamma, U=2\Gamma$). As $U$ is increased, the effects of driving  are diminished and at $U=3\Gamma$ they are completely gone. 

\begin{figure}
\centering
\includegraphics[width=\columnwidth]{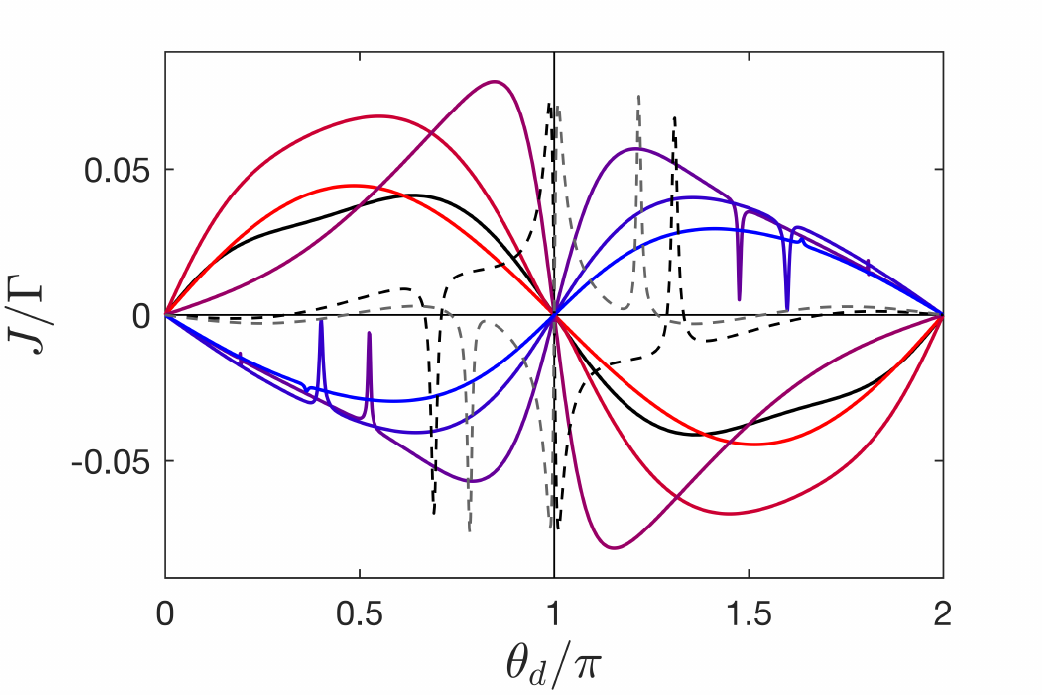}
\caption{Anomalous Josephson current at $\varphi_{sc}=0$ versus driving phase difference, $\theta_{d}$, for interaction strengths ranging from $U=0$ (red) to $U=5\Gamma$ (blue) in steps of $\Gamma$. A marked sign change in current takes place between $U=2.3\Gamma$ (thin black dashed) and $U=2.35\Gamma$ (thin gray dashed). Other parameters are as for the red curve in Fig.~\ref{fig:posj} ($\Omega = 2.2\Gamma$, $\theta_d=\pi/2$, $\varepsilon_d = 0.8\Gamma$, $t_d = 2\Gamma$, $A=0.7\Gamma$, and $n_{max}=9$).}
\label{fig:AppFig2}
\end{figure}

The red curve in Fig.~\ref{fig:posj}, corresponding to $A=0.7\Gamma$, displays a finite anomalous Josephson current at $\varphi_{sc}=0$. In Fig.~\ref{fig:AppFig2}, we use the same parameters to show that this anomalous Josephson current depends strongly on the phase difference of the two drives, $\theta_{d}$, as found also with a finite BCS gap in the red curve of the right panel of Fig.~\ref{fig:phiphi}. Here, however, one observes also a sign change of the anomalous Josephson current, corresponding to a transition from a $0$- to $\pi$-junction behavior in $\theta_{d}$, when increasing the interaction strength. For the chosen parameters, this takes place at a critical interaction strength, $U_{c}\sim\Omega$, but the more detailed parametric dependence of $U_{c}$ is beyond the scope of this paper. 

\begin{figure}
\centering
\includegraphics[width=\columnwidth]{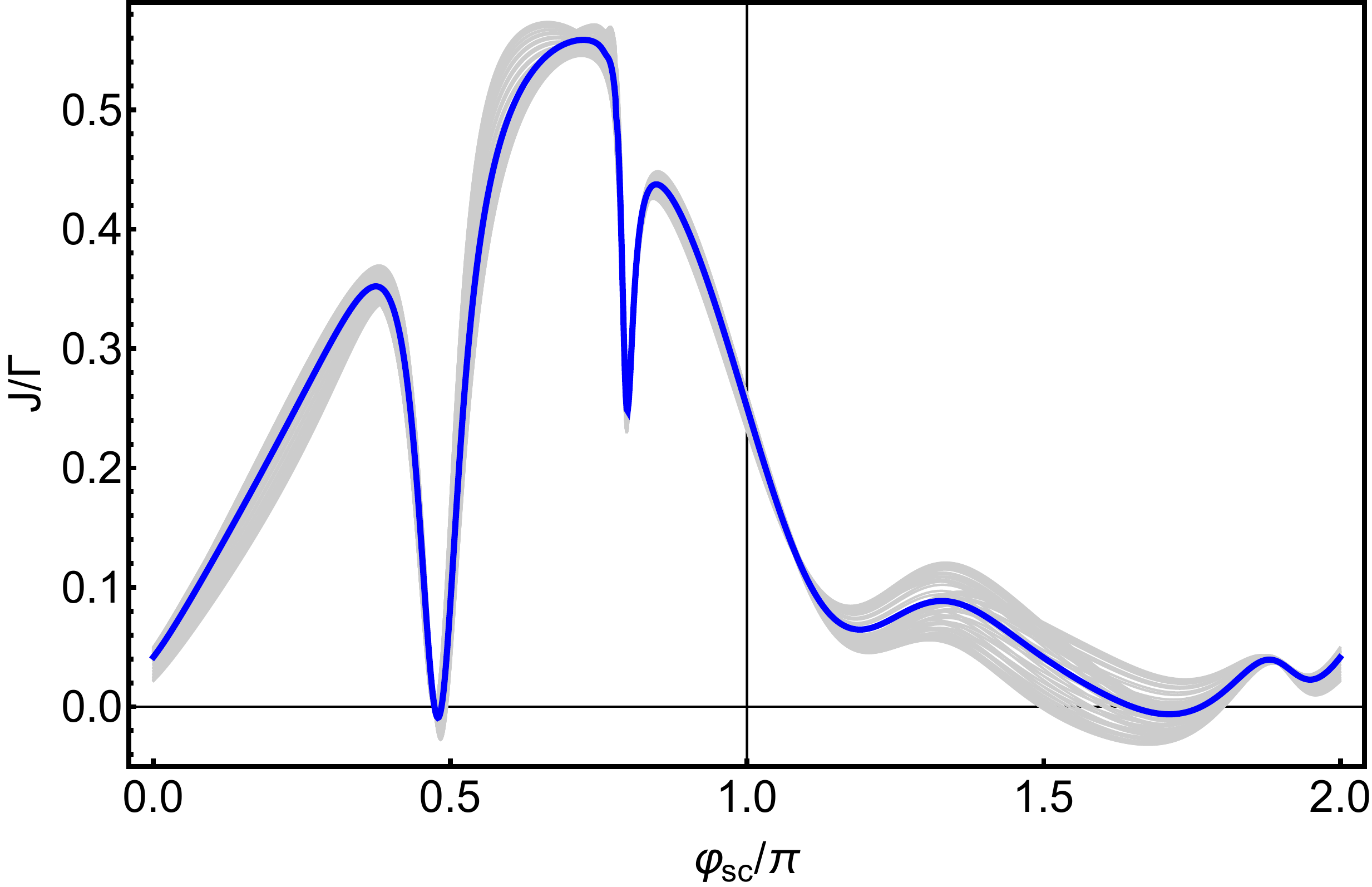}
\caption{Nearly rectified time-averaged current phase relations in the infinite-gap limit for 63 evenly spaced values of $\theta_{R}\in[0,2\pi]$ (gray curves) together with the corresponding $\theta_{R}$-averaged (blue) curve. Parameters are chosen to be $\Omega = 2.1\Gamma$, $\theta_d=\pi/3$, $\varepsilon_d = 0.95\Gamma$, $t_d = 2\Gamma$, $A=0.94\Gamma$, and $U=0.045\Gamma$.}
\label{fig:AppFig3}
\end{figure}

Finally, with Fig.~\ref{fig:AppFig3}, we demonstrate that nearly complete rectification of the time-averaged current is possible also in the infinite-gap limit, where all current is carried by Cooper pairs. Unlike the finite-gap results shown in Figs.~\ref{fig:eda} and~\ref{fig:phiphi}, parameters have been fine tuned so as to make the current positive for all phase differences, $\varphi_{sc}$. 

In Fig.~\ref{fig:AppFig3}, we demonstrate also an explicit dependence of the current on the initial conditions as a spread in curves obtained for different Floquet gauges~\cite{Bukov2015Mar}, corresponding to different values of $\theta_{R}$. This is indicated by a set of some 63 gray curves, corresponding to evenly spaced values of $\theta_{R}$ between $0$ and $2\pi$, which are averaged to obtain the blue curve. A similar spread will be obtained for the curves in Fig.~\ref{fig:AppFig2} (not shown for clarity), whereas in Fig.~\ref{fig:AppFig1}, the driving amplitude is low enough that the results depend only on the phase difference, $\theta_{d}$. This spread increases with driving amplitude and gives a rough indication of sensitivity of the long time average of the Floquet time evolved current on initial conditions, and thereby whether they can be expected to be valid also within a driven steady state.


\subsection{Symmetries of the current}

The time-dependent current and thereby its long-time average obeys a few basic symmetries, which are most easily revealed by reverting to the time-dependent infinite-gap Hamiltonian for the even sector obtained by replacing $\varepsilon_{d}$ by $\varepsilon_{d}(t)$ in $\hat{H}_{e,\infty}$. The corresponding time-dependent infinite gap Hamiltonian, $\hat{H}(\varepsilon_{d},\varphi_{sc},\theta_{L},\theta_{R},t)=\hat{H}_{e,\infty}|_{\varepsilon_{d}\rightarrow\varepsilon_{d}(t)}$ and the current operator obey the transformation properties
\begin{align}
\hat{\mathcal I}\hat{H}(\varphi_{sc},\theta_{L},\theta_{R},U,\varepsilon_{d},t)\hat{\mathcal I}&=\label{eq:hamtrans}\\
&\hspace*{-22mm}\hat{H}(-\varphi_{sc},\theta_{R},\theta_{L},U,\varepsilon_{d},t),\nonumber\\
\hat{\mathcal C}\hat{H}(\varphi_{sc},\theta_{L},\theta_{R},U,\varepsilon_{d},t)\hat{\mathcal C}&=\nonumber\\
&\hspace*{-22mm}\hat{H}(-\varphi_{sc},\theta_{L}+\pi,\theta_{R}+\pi,U,-\varepsilon_{d}-U,t)+\delta\hat{H},\nonumber
\end{align}
and
\begin{align}
\hat{\mathcal I}\hat{I}(\varphi_{sc})\hat{\mathcal I}&=-\hat{I}(-\varphi_{sc}),\\
\hat{\mathcal C}\hat{I}(\varphi_{sc})\hat{\mathcal C}&=-\hat{I}(-\varphi_{sc}),\nonumber
\end{align}
with orthogonal matrices given by,
\begin{align}
\hat{\mathcal I}&=\left(
\begin{array}{cccccc}
1 & 0 & 0 & 0 & 0 & 0  \\
0 & 0 & 1 & 0 & 0 & 0  \\
0 & 1 & 0 & 0 & 0 & 0  \\
0 & 0 & 0 & 1 & 0 & 0  \\
0 & 0 & 0 & 0 & 0 & -1 \! \\
0 & 0 & 0 & 0 & -1 & 0
\end{array}\right)\!\!,\,
\hat{\mathcal C}&=\left(
\begin{array}{cccccc}
0 & 0 & 0 & 1 & 0 & 0  \\
0 & 0 & 1 & 0 & 0 & 0  \\
0 & 1 & 0 & 0 & 0 & 0  \\
1 & 0 & 0 & 0 & 0 & 0  \\
0 & 0 & 0 & 0 & 0 & -1 \! \\
0 & 0 & 0 & 0 & -1 & 0
\end{array}\right)\!,
\end{align}
with $\hat{\mathcal I}$ corresponding to inversion, while $\hat{\mathcal C}$ is related to charge conjugation, but defined here without the complex conjugation operator. The correction term induced by $\hat{\mathcal C}$ has matrix elements
\begin{align}
\delta\hat{H}_{ij}=\left[4\varepsilon_{d}+2U+2A\!\!\sum_{\alpha=L,R}\cos(\theta_{\alpha}+\Omega t)\right]\delta_{ij}.
\end{align}
which merely shifts the diagonal terms, and amounts simply to a multiplicative phase factor between the transformation partner states. From the transformation properties (\ref{eq:hamtrans}), one finds the transformation of a given solution to the time-dependent Schr\"{o}dinger equation to be itself a solution with different parameters, namely:
\begin{align}
\hat{\mathcal I}|\Psi(\varphi_{sc},\theta_{L},\theta_{R},U,\varepsilon_{d},t)\rangle&=\\
&\hspace*{-32mm}|\Psi(-\varphi_{sc},\theta_{R},\theta_{L},U,\varepsilon_{d},t)\rangle,\nonumber\\
\hat{\mathcal C}|\Psi(\varphi_{sc},\theta_{L},\theta_{R},U,\varepsilon_{d},t)\rangle&=\nonumber\\
&\hspace*{-32mm}
e^{-i(\Theta(t)-\Theta(0))}|\Psi(-\varphi_{sc},\theta_{L}+\pi,\theta_{R}+\pi,U,-\varepsilon_{d}-U,t)\rangle,\nonumber
\end{align}
where the common time-dependent phase factor has been introduced as
\begin{align}
\Theta(t)=4(\varepsilon_{d}+U/2)t+4(A/\Omega)\!\!\sum_{\alpha=L,R}\!\!\sin(\theta_{\alpha}+\Omega t).   
\end{align}

\begin{figure}[t]
\centering
\includegraphics[width=\columnwidth]{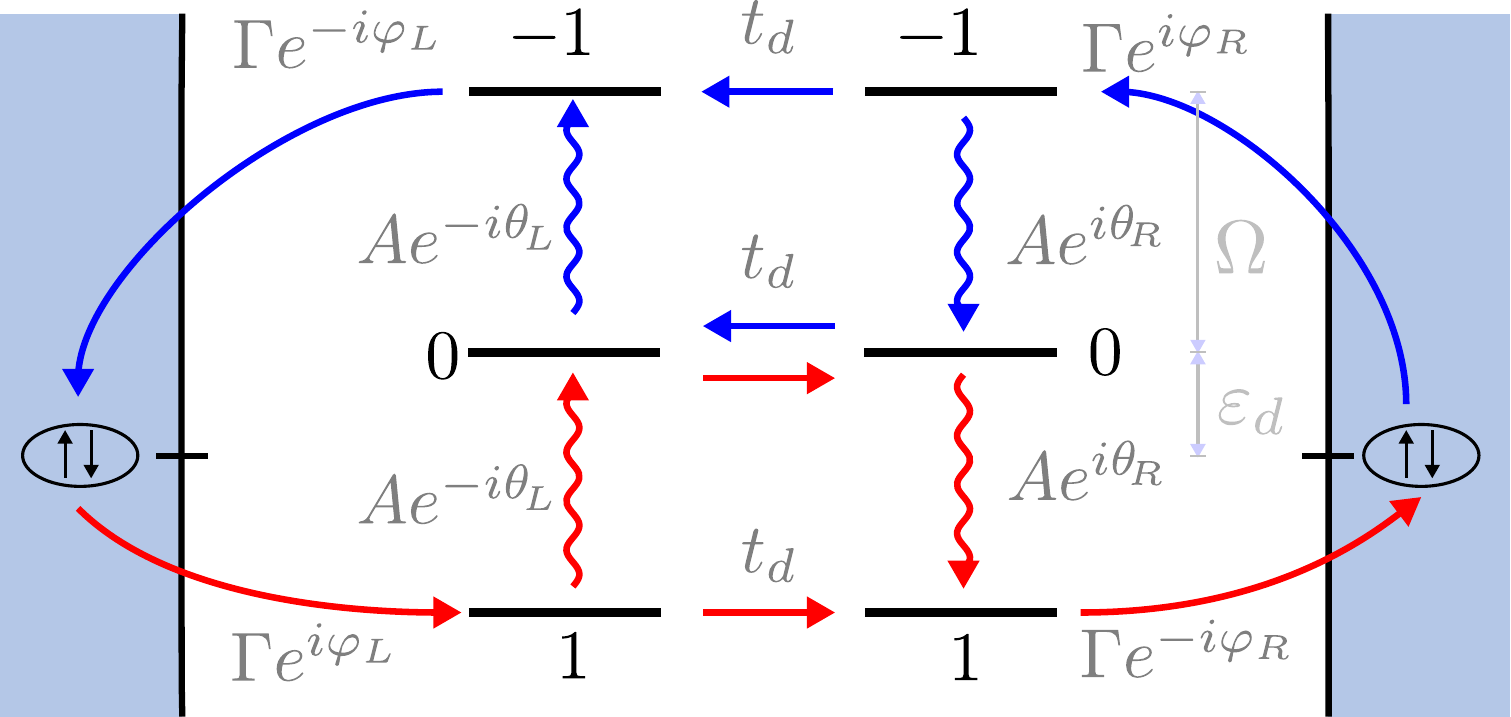}
\caption{Diagram illustrating the lowest order transport processes leading to anomalous Josephson current. The processes are similar to those illustrated in Fig.~\ref{fig:bspt}, but are shown here with two Floquet side bands to each quantum dot level and with current carrying tunnelling paths in both directions, which interfere destructively unless $\varepsilon_{d}\neq 0$, or with interactions included, $\varepsilon_{d}+U/2\neq 0$.}
\label{fig:pump_2}
\end{figure}
Together with the transformation properties of the current operator, this implies that
\begin{align}
&I(-\varphi_{sc},\theta_{R},\theta_{L},U,\varepsilon_d,t)
\nonumber\\
&\hspace*{5mm}=\langle \Psi(\varphi_{sc},\theta_{L},\theta_{R},U,\varepsilon_{d},t)|\hat{\mathcal I}\hat{I}(-\varphi_{sc})\hat{\mathcal I}\nonumber\\
&\hspace*{45mm}\times|\Psi(\varphi_{sc},\theta_{L},\theta_{R},U,\varepsilon_{d},t)\rangle\nonumber\\
&\hspace*{5mm}=-I(\varphi_{sc},\theta_{L},\theta_{R},U,\varepsilon_d,t),
\end{align}
and
\begin{align}
&I(-\varphi_{sc},\theta_{L}+\pi,\theta_{R}+\pi,U,-\varepsilon_{d}-U,t)
\nonumber\\
&\hspace*{5mm}=\langle\Psi(\varphi_{sc},\theta_{L},\theta_{R},U,\varepsilon_{d},t)|\hat{\mathcal C}\hat{I}(-\varphi_{sc})\hat{\mathcal C}\nonumber\\
&\hspace*{45mm}\times|\Psi(\varphi_{sc},\theta_{L},\theta_{R},U,\varepsilon_{d},t)\rangle\nonumber\\
&\hspace*{5mm}=-I(\varphi_{sc},\theta_{L},\theta_{R},U,\varepsilon_d,t).
\end{align} 
From these instantaneous symmetries one may infer the symmetries~\eqref{eq:1s} and~\eqref{eq:2s} of the time-averaged currents, 
\begin{align}
J(\varphi_{sc},\theta_d,U,\varepsilon_d)
&=- J(-\varphi_{sc},\theta_d,U,-\varepsilon_d-U),\label{eq:phsym}\\
&=-J(-\varphi_{sc},-\theta_d,U,\varepsilon_d),\label{eq:Isym}\\
&=J(\varphi_{sc},-\theta_d,-\varepsilon_d-U)\label{eq:sym3},
\end{align}
which are observed also in the non-interacting finite-gap numerical results shown in Figs.~\ref{fig:eda} and~\ref{fig:phiphi}. The inversion symmetry relation (\ref{eq:Isym}) alone dictates that the anomalous Josephson current must vanish at $\theta_{d}=\pi=2\pi-\pi$, as observed in Fig.~\ref{fig:AppFig2}. The particle-hole symmetry relation (\ref{eq:phsym}), and thereby (\ref{eq:sym3}), holds only when the average phase of the two drives plays no role, i.e. when either the driving amplitude is sufficiently small or when all transients have been erased by relaxation via the quasiparticle continuum available for finite BCS gaps or weak tunnelling to normal metals as modelled by $\Gamma_{m}$ in the NESS Floquet Keldsyh Green function method employed in the main text. 

From these symmetries, the anomalous Josephson current at $\varphi_{sc}=0$ is seen to satisfy the symmetries 
\begin{align}
J(0,\theta_d,U,\varepsilon_d)=&-J(0,-\theta_d,U,\varepsilon_d)\\
=&-J(0,\theta_d,U,-\varepsilon_d-U).   
\end{align}
This implies that the anomalous Josephson current must vanish for quantum dots tuned to the particle-hole symmetric point, $\varepsilon_{d}=-U/2$. Within the Floquet picture, this vanishing of the anomalous Josephson current at the particle-hole symmetric point can be understood as a destructive interference between paths through respectively positive, and negative Floquet sidebands. This is illustrated in Fig.~\ref{fig:pump_2}, in which the blue and red paths need to be off-set from particle-hole symmetry in order not to interfere destructively.

\bibliographystyle{apsrev4-1}
\bibliography{dqd_floquet_bib.bib}

\end{document}